\documentclass[sigconf]{acmart}

\copyrightyear{2025}
\acmYear{2025}
\setcopyright{acmlicensed}\acmConference[KDD '25]{Proceedings of the 31st ACM SIGKDD Conference on Knowledge Discovery and Data Mining V.1}{August 3--7, 2025}{Toronto, ON, Canada}
\acmBooktitle{Proceedings of the 31st ACM SIGKDD Conference on Knowledge Discovery and Data Mining V.1 (KDD '25), August 3--7, 2025, Toronto, ON, Canada}
\acmDOI{10.1145/3690624.3709422}
\acmISBN{979-8-4007-1245-6/25/08}

\usepackage{amsmath}
\usepackage{algorithmic}
\usepackage{graphicx}
\usepackage{textcomp}
\usepackage{xcolor}
\usepackage{multicol}
\usepackage{amsthm}
\usepackage[linesnumbered,ruled]{algorithm2e}
\usepackage{caption}
\usepackage{subfigure}
\usepackage{url}
\usepackage{multirow}
\usepackage{float}
\usepackage{enumitem}

\AtBeginDocument{%
  }

\begin{document}

\title{Multi-period Learning for Financial Time Series Forecasting}


\author{Xu Zhang}
\authornote{This work was conducted when Xu Zhang was interning at Ant Group. This work was supported by Ant Group Research Intern Program.}

\affiliation{%
  \institution{School of Computer Science\\ Fudan University}
  \city{Shanghai}
  \country{China}
}
\email{xuzhang22@m.fudan.edu.cn}

\author{Zhengang Huang}
\affiliation{%
  \institution{Ant Group}
  \city{Shanghai}
  \country{China}
}
\email{huangzhengang.hzg@antgroup.com}

\author{Yunzhi Wu}
\affiliation{%
  \institution{School of Computer Science\\ Fudan University}
  \city{Shanghai}
  \country{China}
}
\email{yzwu22@m.fudan.edu.cn}

\author{Xun Lu}
\affiliation{%
  \institution{Ant Group}
  \city{Shanghai}
  \country{China}
}
\email{hilber.lx@antgroup.com}

\author{Erpeng Qi}
\affiliation{%
  \institution{Ant Group}
  \city{Shanghai}
  \country{China}
}
\email{erpeng.qep@antgroup.com}

\author{Yunkai Chen}
\affiliation{%
  \institution{Ant Group}
  \city{Shanghai}
  \country{China}
}
\email{chenyunkai.cyk@antgroup.com}

\author{Zhongya Xue}
\affiliation{%
  \institution{Ant Group}
  \city{Shanghai}
  \country{China}
}
\email{zhongya.xzy@antgroup.com}

\author{Qitong Wang}

\authornote{Corresponding author.}
\affiliation{%
  \institution{Universite Paris Cite}
  \city{Paris}
  \country{France}
}
\email{qitong.wang@u-paris.fr}

\author{Peng Wang}
\affiliation{%
  \institution{School of Computer Science\\ Fudan University}
  \city{Shanghai}
  \country{China}
}
\email{pengwang5@fudan.edu.cn}

\author{Wei Wang}
\affiliation{%
  \institution{School of Computer Science\\ Fudan University}
  \city{Shanghai}
  \country{China}
}
\email{weiwang1@fudan.edu.cn}

\renewcommand{\shortauthors}{Xu Zhang, et al.}

\begin{abstract}
Time series forecasting is important in finance domain. Financial time series (TS) patterns are influenced by both short-term public opinions and medium-/long-term policy and market trends. Hence, processing multi-period inputs becomes crucial for accurate financial time series forecasting (TSF).
However, current TSF models either use only single-period input, or lack customized designs for addressing multi-period characteristics. In this paper, we propose a Multi-period Learning Framework (MLF) to enhance financial TSF performance.  
MLF considers both TSF's accuracy and efficiency requirements. Specifically, we design three new modules to better integrate the multi-period inputs for improving accuracy: (i) Inter-period Redundancy Filtering (IRF), that removes the information redundancy between periods for accurate self-attention modeling, (ii) Learnable Weighted-average Integration (LWI), that effectively integrates multi-period forecasts, (iii) Multi-period self-Adaptive Patching (MAP), that mitigates the bias towards certain periods by setting the same number of patches across all periods. Furthermore, we propose a Patch Squeeze module to reduce the number of patches 
in self-attention modeling for maximized efficiency. 
MLF incorporates multiple inputs with varying lengths (periods) to achieve better accuracy and reduces the costs of selecting input lengths during training.
The codes and datasets are available at {\url{https://github.com/Meteor-Stars/MLF}}.
\end{abstract}

\ccsdesc[500]{Information systems~Spatial-temporal systems}
\ccsdesc[500]{Computing methodologies~Artificial intelligence}

\keywords{time series forecasting; deep learning; financial sales; multi-periods; multi-scales; spatio-temporal data mining}

\maketitle

\vspace{-0.25cm}
\section{Introduction}
Time series forecasting (TSF) is a critical task in the finance industry~\cite{sezer2020financial,DBLP:conf/esann/KrollnerVF10,DBLP:journals/widm/ZhangSI24,DBLP:journals/ijon/TangSZYHJTL22}.
For example, Alipay is a well-known App
that offers convenient digital payment and investment services ~\cite{zhang2024self,DBLP:conf/dasfaa/ZhangCZL18,DBLP:conf/kdd/ZangHC0ZZZ23,DBLP:conf/ijcai/ChenZJFTZL19}. 
Fund sales forecasting is used in financial services of Alipay App to manage the inventory of different fund products. 
Accurate inventory management is not only crucial for ensuring sufficient availability of various fund products on the Alipay App, but also for aiding fund institutions in investment preparation and risk management. 
Moreover, TSF models need to be daily retrained and inferred to accommodate new data for the most accurate fund sales forecasting. 
Therefore, not only accuracy, but also the efficiency of the TSF model are crucial for real-world financial applications.

One of the key characteristics of financial TS is that their temporal patterns are influenced by different information from multiple periods of historical data  (i.e., TS with multiple lengths) ~\cite{DBLP:conf/ccbd/ChenCHHC16,DBLP:phd/basesearch/Zeng08,DBLP:journals/complexity/GuX21,DBLP:journals/tnn/TranIKG19}. 
For instance, the sales of fund products 
are affected not only by short-term public opinions and the company's operational conditions, but also by medium-/long-term policies and market trends. 
We also found this phenomenon in our experiments. 
In the scenario of fund sales forecasting, Table~\ref{tab:period_input_comp} indicates that there is no single period input that can provide the best predictions.
Specifically, 33.53\% samples are best forecasted using the short-period input (PTST$_{5}$), 27.35\% using medium-period input (PTST$_{10}$) and 39.12\% using long-period input (PTST$_{30}$), respectively.
These uniform ratios suggest that choosing a single fixed period will result in suboptimal forecasting accuracy. 
For this situation, a simple solution might be to select appropriate input lengths for different forecasting legnths during training, but this leads to high overhead and may not be accurate enough.

 \begin{table} [bt] 
    \centering
    \vspace{-0.2cm}
    \caption{MSE (Mean Squared Error; lower is better) of a single-step TSF under various input lengths $n$ (representing different periods) on the Alipay Fund dataset. 
    PTST$_{5}$ refers to a PatchTST~\cite{nie2022time_patchformer} model trained on inputs of length $n=5$. 
    $\kappa=mean ((X_h-X_f)^2)$ quantifies the consistency between historical windows $X^h$ (30 time steps) and the forecasted value $X^f$. 
    A higher $\kappa$ indicates sharper pattern fluctuations.
    }
    \vspace{-0.1cm}
    \label{tab:period_input_comp}
    { \small
\begin{tabular}{c|ccc|c}
    \hline
    \multirow{2}{*}{\shortstack{Data Subset Ratios}}  &  \multicolumn{3}{c|}{MSE} &\multicolumn{1}{c}{Distribution} \\
    &PTST$_{5}$ &PTST$_{10}$&PTST$_{30}$&Consistency $\kappa$\\
     \midrule[0.5pt]
     \multirow{1}{*}{33.53\%, best by PTST$_{5}$} & \textbf{49.28} &54.85 &63.31&60.64 \\
    \midrule[0.5pt]
     \multirow{1}{*}{27.35\%, best by PTST$_{10}$} &12.34 & \textbf{8.92}  &13.57&22.46\\
    \midrule[0.5pt]
     \multirow{1}{*}{39.12\%, best by PTST$_{30}$} &14.56 &12.23 & \textbf{7.34}  &14.82\\
    \midrule[0.5pt]
\end{tabular}
    } 
\vspace{-0.2cm}
\end{table}

Further analyzing the temporal pattern, we found 33.53\% samples, best predicted by short-period input, tend to have sharply fluctuating future values.
This is evidenced by the highest $\kappa$=60.64.
In contrast, the remaining 27.35\% and 39.12\% samples, best predicted by medium- and long-period inputs, have relatively smooth future values 
as they show lower $\kappa$=22.46 and 14.82, respectively.  
This can be interpreted in the financial domain that short-term hot public topics lead to urgent fund transactions, 
while medium-/long-term global policies and market trends have a bigger impact on the steady trend of fund sales. 
Hence, short-period inputs are tightly correlated with future fluctuations, while medium- and long-period inputs are more 
associated with future steady trends. 
This explains why different periods work best at predicting different samples in Table~\ref{tab:period_input_comp}.
In fact, future values 
can consist of both sharp fluctuations and smooth trends. 
In this case, even selecting appropriate input lengths in advance for different prediction lengths cannot address the issue. 
Hence, considering the multi-period inputs simultaneously is crucial for accurate financial TSF.

\begin{figure}[bt]
\vspace{-0.2cm}
\centerline{\includegraphics[width=\linewidth]{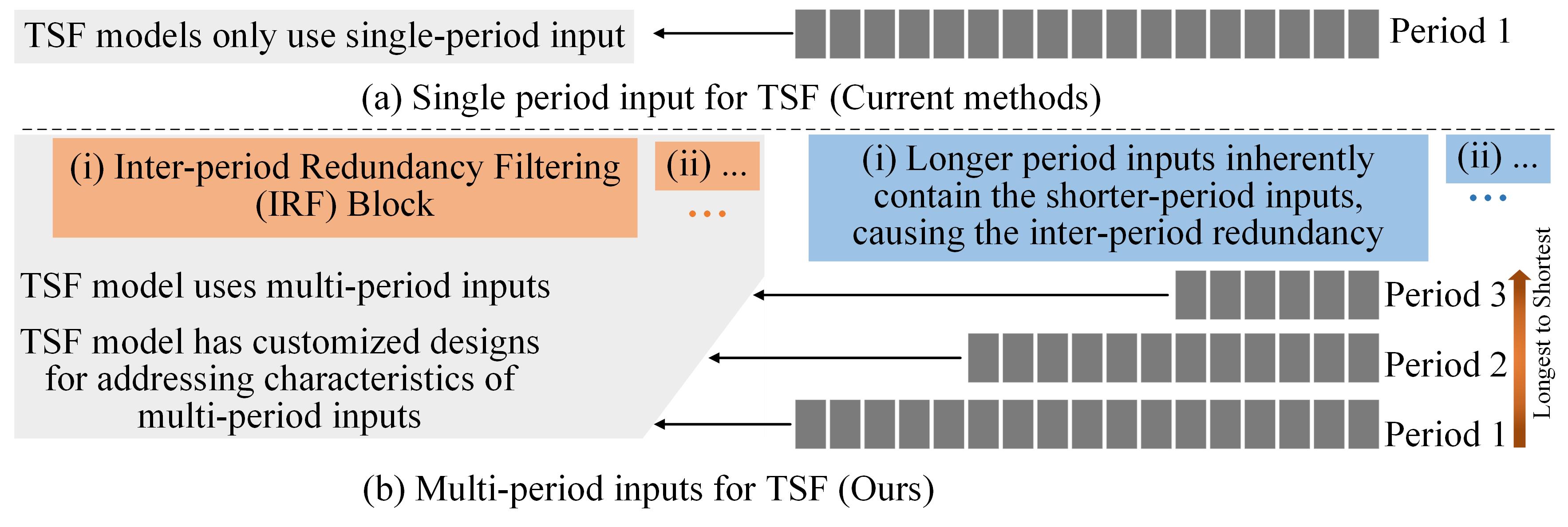} }
\vspace{-0.2cm}
\caption{Current TSF scheme (a) and ours (b). The blue rectangle highlights the characteristics of the multi-period inputs.}
\vspace{-0.2cm}
\label{fig:multi_scale_inputs_0624}
\end{figure}

Recently, several architectures have been proposed for TSF task, including transformer-based Scaleformer~\cite{shabani2022scaleformer}, PatchTST~\cite{nie2022time_patchformer} and Pathformer~\cite{chen2024pathformer}, as well as linear model-based NHits~\cite{challu2022n_Nhits}, TSmixer~\cite{chen2023tsmixer} and TiDE~\cite{das2023long}.
These models show promising accuracy in the TSF task, but their designs only consider the single-period input (TS with a fixed length), as shown in Figure~\ref{fig:multi_scale_inputs_0624}(a). 
On one hand, this leads to the need to select appropriate input lengths for different prediction lengths, resulting in high training overhead. On the other hand, as analyzed above, future values may contain both sharp fluctuations and stable trends, which requires joint predictions using inputs of different lengths. 
Although FiLM~\cite{zhou2022film} considers 
the multi-period inputs, 
it just linearly integrates multi-period outputs
without any specific architecture designs for multi-period characteristics. 
We discuss one multi-period property in the right of Figure~\ref{fig:multi_scale_inputs_0624}(b).
As longer-period inputs inherently contain shorter-period inputs, there is information redundancy among different periods.
If not well processed, this inter-period redundancy may have adverse effects on model training.
For example, it can cause self-attention to overfocus on the information of repetitive parts among different periods, leaving the other parts underutilized. 
Hence,  in our TSF scheme, 
TSF model not only takes multi-period inputs, but also embrace customized designs to address the multi-period characteristics, as shown in the left of Figure~\ref{fig:multi_scale_inputs_0624}(b). 
To our knowledge, such multi-period based TSF model is rarely explored.

In this paper, we introduce MLF (Multi-period Learning Framework) for accurate financial TSF.
MLF is designed to effectively utilize the diverse information across 
multi-period inputs. 
However, it is nontrivial to integrate multi-period inputs, which
presents challenges in accuracy and efficiency.
First, how to address the characteristics of multi-period inputs to effectively integrate multi-period information for forecasting is a key challenge (\textit{integration challenge}). 
Second, compared to single-period inputs in Figure~\ref{fig:multi_scale_inputs_0624}(a), 
multi-period inputs in Figure~\ref{fig:multi_scale_inputs_0624}(b) 
will significantly increase the computational and memory usage of the model. 
Therefore, 
how to enhance the efficiency of multi-period based models while maintaining improved accuracy is another challenge (\textit{efficiency challenge}).
To address the \textit{integration challenge}, we introduce three new designs into the transformer architecture: (i) Inter-period Redundancy Filtering (IRF) block to eliminate the inter-period information redundancy,
(ii) Learnable Weighted-average Integration (LWI) module to effectively integrate the final multi-period forecasts,
(iii) Multi-period self-Adaptive Patching (MAP) module to set the same number of patches for all periods. 
To tackle the \textit{efficiency challenge}, we propose a Patch-Squeeze module to reduce input length by removing redundancy within each single period (i.e., intra-period redundancy). 
Specifically, the motivation of integration design (i) is that redundancy naturally exists between different periods. 
Due to similar segments having higher correlations, this redundancy may cause the self-attention to excessively focus on repetitive segments and can't effectively utilize information in non-repetitive segments among periods for more accurate multi-period integration.
We propose an Inter-period Redundancy Filtering (IRF) block to address this issue. 
For integration design (ii), 
before reaching the Learnable Weighted-average Integration (LWI) module, the model's output is still multi-period forecasts, so we need to make the final integration of them. 
Considering different future positions are best forecasted by different periods,
simply averaging may lead to inaccurate integration. 
Hence, we propose LWI to adaptively assign weights to different period forecasts, giving higher weights to more accurate predictions and vice versa. 
Accordingly, LWI can improve the integration accuracy of multi-period forecasts.
The design intuition for integration design (iii) is that the model may have a bias towards certain periods due to they have varying numbers of patches. 
This may prevent the model from fully utilizing information of all periods, thereby reducing integration accuracy. 
Hence, we propose Multi-period self-Adaptive Patching (MAP) to set the same number of patches for all periods to address this issue.

Finally, employing multi-period inputs will significantly increase time and memory costs, posing challenges for scaling to large datasets and deploying in a production system.
To address the \textit{efficiency challenge}, we design a Patch-Squeeze module to reduce the number of patches within each period.
Patch-Squeeze is based on a key observation, originally from the TS self-supervised learning tasks, that a few numbers of patches are sufficient in reconstructing a much larger set of masked patches~\cite{shao2022pre}.
This can be interpreted as there exists intra-period information redundancy.
Hence, reducing intra-period redundancy, i.e., reducing the number of patches within each period, can improve model efficiency without compromising accuracy.

In summary, compared to existing methods, MLF processes multiple inputs of different lengths to enhance prediction accuracy, considering both accuracy and efficiency. Our contributions are as follows:
\begin{enumerate}[noitemsep, topsep=0pt, wide=\parindent]
\item We introduce MLF, a new Multi-period Learning Framework for financial TSF. MLF incorporates multiple inputs with varying lengths (periods) to achieve better accuracy, and one benefit is reducing the costs of selecting input lengths for different prediction lengths during training.

\item We propose three new architecture designs to better integrate the multi-period inputs, namely, Multi-period self-Adaptive Patching (MAP) modules, Inter-period Redundancy Filtering (IRF) block, and Learnable Weighted-average Integration (LWI).

\item We further propose a simple but effective Patch-Squeeze module for time dimensionality reduction, which provides high efficiency and maintain expected TSF accuracy while addressing multi-period inputs with varying lengths.

\item  We introduce a new financial dataset of fund sales collected from Ant Fortune and the Alipay App, expanding the public dataset repository and enabling a more comprehensive evaluation of TSF models.

\item Our experimental results indicate that MLF outperforms advanced TSF baselines in terms of accuracy and efficiency on both collected fund and public datasets. 
The deployment results on the Alipay App further confirms MLF's effectiveness. 
\end{enumerate}

\section{Related Work and Preliminary}
\label{sec:related_work}
\subsection{Time Series Forecasting (TSF)}
\subsubsection{Single-period based TSF}
Numerous deep learning models have been proposed by using single-period. 
RNNs~\cite{jordan1997serial,elman1990finding,hochreiter1997long,cho2014learning} use recurrent structures to capture short-term temporal dependency. 
Meanwhile, temporal convolutional networks (TCN)~\cite{sen2019think,GraphWave,DBLP:conf/kdd/WuPL0CZ20,DBLP:conf/iclr/LiYS018} employ causal and dilated convolutions for parallel computation, effectively capturing temporal dependencies in the TSF task.

The transformer~\cite{vaswani2017attention} has been extensively adapted for long-term TSF task~\cite{wu2021autoformer,zhou2022fedformer}.
Pyraformer~\cite{liu2021pyraformer} employs a multi-pass downsampling operation to capture temporal dependencies at various granularities. Scaleformer~\cite{shabani2022scaleformer} further enhances Pyraformer by iteratively refining forecasts at finer scales. 
Instead of using multi-pass downsampling, Pathformer~\cite{chen2024pathformer} utilizes different patch sizes to construct multiple time series and designs an adaptive pathway for improving the performance of TSF. 
However, linear models like DLinear~\cite{zeng2023transformers_linear}, TSMixer~\cite{chen2023tsmixer}, N-BEATS~\cite{oreshkin2019n_N-BEATS} and NHits~\cite{challu2022n_Nhits} have emerged to challenge the effectiveness of transformer-based methods. PatchTST~\cite{nie2022time_patchformer} addresses this by splitting input series into patches and learning self-attention at the patch level, achieving advanced performance in long-term TSF tasks. 
This illustrates that transformers retain significant potential for TSF when appropriately tailored to the task.

Recently, Large Language Models (LLMs) have shown effectiveness in TSF tasks, particularly in scenarios requiring few training examples (few-shot learning)~\cite{jin2023time, zhou2023one}. Nonetheless, researchers also express worry about using LLMs for TSF~\cite{tan2024language}.

\vspace{-0.1cm}
\subsubsection{Multi-period based TSF} 
To our knowledge, few TSF models are proposed to utilize multi-period inputs simultaneously to obtain more accurate TSF. 
FiLM~\cite{zhou2022film} is a multi-period based method, but it just linearly integrates multi-period outputs for TSF. 
Although it shows accuracy improvement for using multi-period inputs, it lacks customized designs to effectively handle their characteristics such as redundancy between periods, varying prediction effectiveness, and varying sequence lengths among periods, which suggests an opportunity for further research. 

 \begin{figure}[h]
\centerline{\includegraphics[width=\linewidth]{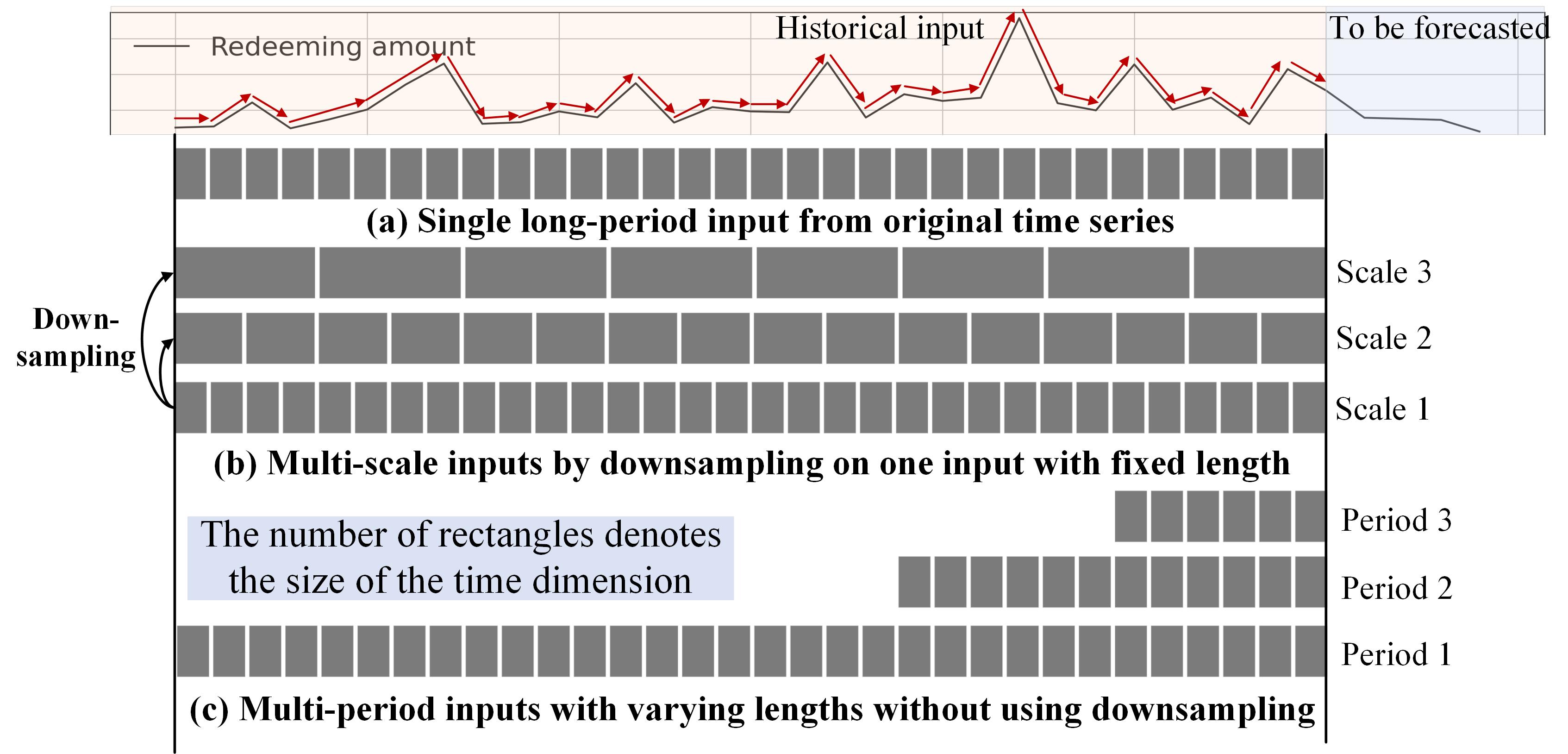} }
\vspace{-0.2cm}
\caption{Comparison between multi-period inputs in subfigure (c) and multi-scale inputs in subfigure (b).}
\vspace{-0.2cm}
\label{fig:Comparison_multiscale_multiperiod}
\end{figure}

\begin{figure*}[h]
\vspace{-0.2cm}
\centerline{\includegraphics[width=0.9\linewidth]{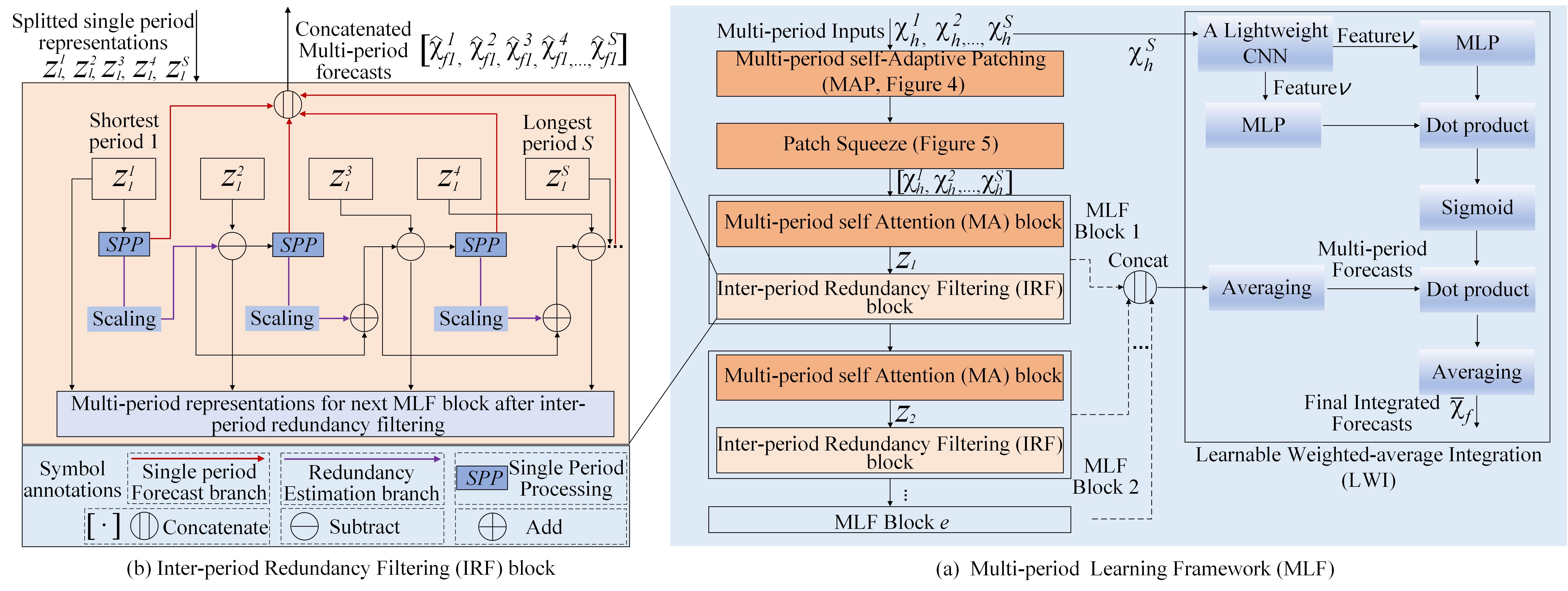} }
\vspace{-0.4cm}
\caption{MLF and corresponding components.}
\label{fig:model}
\vspace{-0.3cm}
\label{fig:frame_work}
\end{figure*}
\subsubsection{Difference between multi-period inputs and multi-scale inputs} 

In this paper, multi-period inputs refer to multiple original time series windows with varying input lengths, as shown in Figure~\ref{fig:Comparison_multiscale_multiperiod}(c). 
This is different from the multi-scale inputs in Pyraformer~\cite{liu2021pyraformer} and Scaleformer~\cite{shabani2022scaleformer}, which are obtained by downsampling from the same fixed input length (Figure~\ref{fig:Comparison_multiscale_multiperiod}(b)). 
In extensive experiments, we observe that different input lengths have a significant impact on prediction accuracy. However, selecting appropriate input lengths is a crucial challenge affecting time series forecasting.

To address this, we propose MLF to extract the semantic information of short-medium-long-term individually using sequences with varying lengths, to avoid models failing to learn the different semantics under only long-term inputs, e.g., the prediction error of Pathformer and Scaleformer using long-term sequence inputs is higher than that of short-term ones. 
In contrast, by directly inputting multiple windows with varying lengths, MLF can achieve good prediction accuracy, reducing the costs of selecting input windows during training.

\vspace{-0.2cm}
\subsection{Problem Definition}
\label{sec:Problem_Definition}

Given a historical multivariate TS instance $\mathcal{X}_h = [x_1, x_2, \ldots, x_n] \in \mathbb{R}^{n \times c}$ with length $n$, the TSF task aims to predict the future $m$ steps $\mathcal{X}_f=[x_{n+1},x_{n+2},...,x_{n+m}] \in \mathbb {R}^{m \times c}$ for all $c$ variables (for simplicity, we will omit $c$ in the following sections). 

In this paper, we focus on designing a Multi-period Learning Framework for TSF. 
The multi-period inputs $\mathcal{X}_h^*$ are defined as $[\mathcal{X}_h^1, \mathcal{X}_h^2, \ldots, \mathcal{X}_h^S]$ where $s$ ($1 \leq s \leq S$) denotes the period index. 
The multi-period input consists of multiple time series with different lengths $n_s$. 
A larger $s$ indicates a longer period, and $n_S$ represents the length of the longest period.

\vspace{-0.1cm}

\section{Multi-period Learning Framework (MLF)}

We illustrate MLF in Figure~\ref{fig:frame_work}. 
First, the multi-period input passes through the Multi-period self-Adaptive Patching (MAP) module.
The input of each period is converted into the same number of patches and then linearly embedded. 
Second, each period's patch embeddings are sent to the Patch Squeeze module.
It reduces the number of patches by reducing the information redundancy within each period.
Third, the squeezed period patch embeddings are fed into the Multi-period self-Attention (MA) and Inter-period Redundancy Filtering (IRF) blocks. 
IRF removes information redundancy between periods for more accurate MA modeling. 
MA and IRF blocks form a MLF block.
By stacking MLF blocks, MLF progressively removes inter-period redundancy and achieves better modeling of multi-period input. 
Finally, the Learnable Weighted-average Integration (LWI) module integrates multi-period forecasts from all blocks to generate the final prediction.
We will detail each design.

\vspace{-0.2cm}
\subsection{Multi-period self-Adaptive Patching (MAP)}
Following the convention~\cite{nie2022time_patchformer}, we split the multi-period input $\mathcal{X}_h^*$ into patches instead of individual points as input units.

 \vspace{-0.2cm}
 \begin{equation}
\label{equ:patch}
\mathcal{X}_p^{s}=Patch(\mathcal{X}_h^s,L,K),
 \vspace{-0.1cm}
\end{equation}
 \vspace{-0.1cm}
 \begin{equation}
\label{equ:patch_or_NS}
N^s=\lfloor (n^s-L)/K \rfloor +2,
\end{equation}

\noindent where $Patch(\cdot)$ represents the patching function. 
$L$ and $K$ denote the patch length and stride of the $Patch(\cdot)$ function, respectively. 
The parameter $K$ determines the non-overlapping region between consecutive patches. Following~\cite{nie2022time_patchformer}, we pad $K$ repeated numbers of the last value to the end of the original sequence before patching. 
In Eq.~\ref{equ:patch_or_NS}, $\lfloor \cdot \rfloor$ denotes rounding down. 
$Patch(\cdot)$ transforms a time series into multiple subsequences (patches). 
$\mathcal{X}_p^{s}$ denotes the patched period $\mathcal{X}_h^s$. 
Each $\mathcal{X}_p^{s} \in \mathbb {R}^{L \times N^s}$, where $N^s$ represents the number of patches in $\mathcal{X}_h^s$.

\begin{figure}[bt]
\vspace{-0.2cm}
\centerline{\includegraphics[width=0.8\linewidth]{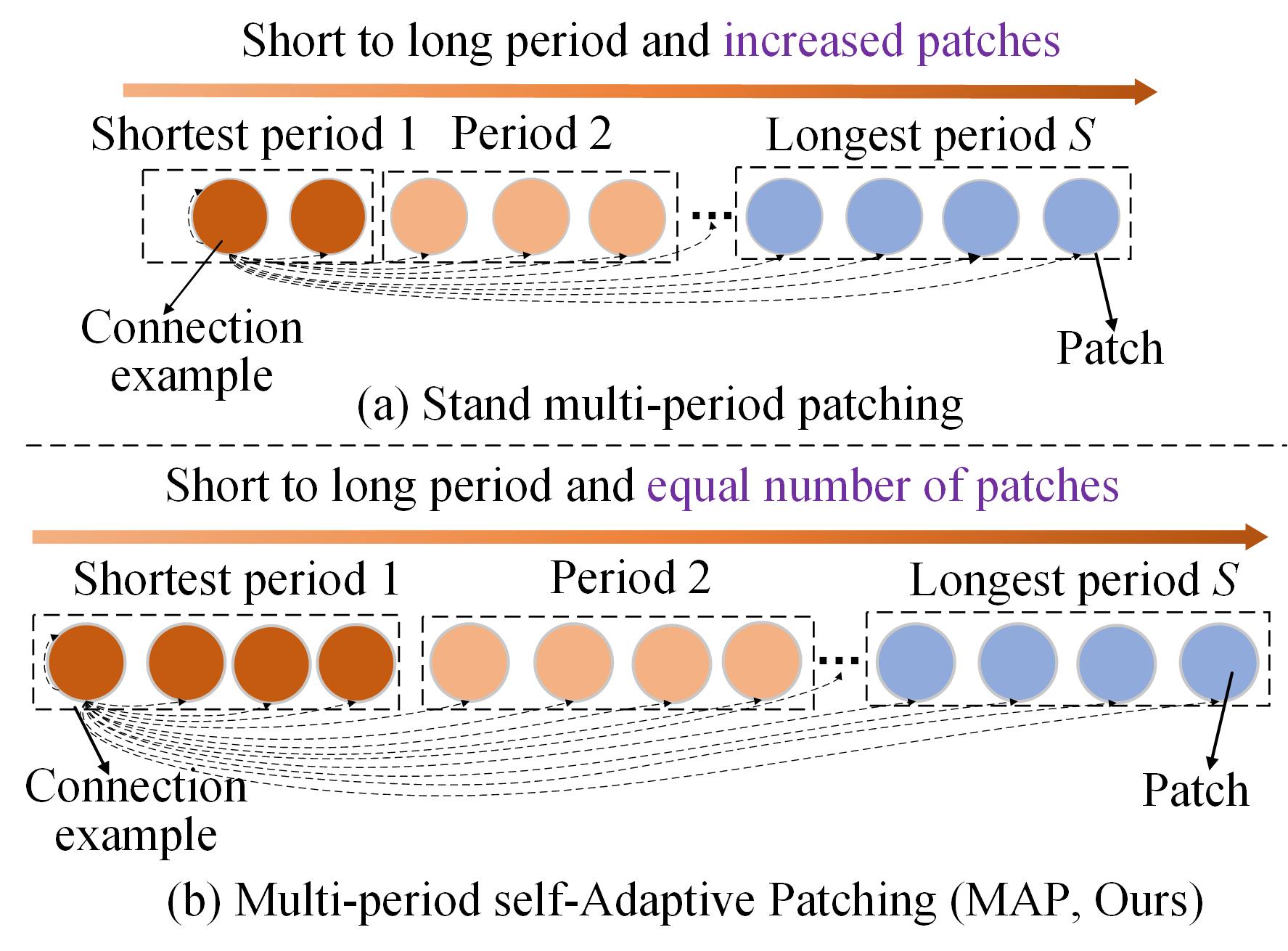} }
\vspace{-0.2cm}
\caption{
Illustration of the standard multi-period patching (using fixed patch lengths) and self-adaptive multi-period patching (using self-adaptively varied patch lengths).}
\vspace{-0.3cm}
\label{fig:adaptive_patching}
\end{figure}

Eq.~\ref{equ:patch} to Eq.~\ref{equ:patch_or_NS} depict standard multi-period patching, where different periods $\mathcal{X}_h^s$ are patched based on fixed values of $L$ and $K$. 
As illustrated in Figure~\ref{fig:adaptive_patching}(a), this approach results in an 
increase in the number of patches from short-term to long-term periods. 
This disparity can lead to unequal treatment of different periods by the model, as shorter-term period inputs 
may not receive adequate utilization %
during optimization due to fewer patches. 

To address this issue of unequal treatment, we introduce Multi-period self-Adaptive Patching (MAP) module. 
Illustrated in Figure~\ref{fig:adaptive_patching}(b), MAP ensures each period $\mathcal{X}_h^s$ contains an equal number of patches by self-adaptively adjusting patch lengths and strides. 
Specifically, we set $L^s=\alpha\cdot K^s$ (where $\alpha$ is fixed at 2 in this paper) and maintain a fixed number of patches $\widetilde{N}$ across all periods. 
Substituting this into Eq.~\ref{equ:patch_or_NS} and expanding, we obtain:
\begin{align}
\widetilde{N} &= (n^s - L)/K + 2 \\
&= (n^s - \alpha\cdot K)/K + 2
\end{align}

After further expanding the formula, we can derive the self-adaptive $L^s$ and $K^s$ for different periods: 
 \begin{equation}
 \vspace{-0.1cm}
\begin{split}
\widetilde{N}\cdot K-\alpha\cdot K &= n^s-\alpha\cdot K \\
\end{split}
\end{equation}
 \begin{equation}
\label{equ:patch_ada_L_K}
K^s=\lfloor n^s/\widetilde{N} \rfloor, \ \ \ L^s=\alpha\cdot K^s 
\end{equation}

Self-adaptive $L^s$ and $K^s$ ensure each period $\mathcal{X}_h^s$ contains an equal number of patches, allowing the model to evenly consider the temporal information across different periods. 
Substituting $L^s$ and $K^s$ into Eq.~\ref{equ:patch} yields the self-adaptive patched $\mathcal{X}_p^{s}$. 

Before feeding the patches of each period into the transformer encoder, they are projected into a $D$-dimensional embedding space with a learnable additive position encoding~\cite{vaswani2017attention,nie2022time_patchformer}. 
The embedded representation of $\mathcal{X}_p^{s}$ is denoted as $\mathcal{X}_d^{s}$ and can be formulated as:
 \begin{equation}
\label{equ:linear_proj_pos}
\mathcal{X}_d^{s}= W{_{p}^{s}}\mathcal{X}_p^{s}+W_{pos}^{s}
\end{equation}
where each $W_p^s \in \mathbb {R}^{D \times L^s}$ and $W_{pos}^s \in \mathbb {R}^{D \times N^s}$.

\subsection{Patch Squeeze}
Studies of self-supervised learning for time series ~\cite{shao2022pre} suggest that there is information redundancy within periods.
That is, a small subset of patches can effectively capture the global temporal patterns of the entire period. 
Motivated by this insight, we propose a patch squeeze module to distill essential information from the original time series data into a reduced number of patches. 
This module not only substantially reduces both time complexity and memory usage, but also shows an insignificant 
impact on accuracy (empirically verified in Section~\ref{sec:exp-efficiency} and Section~\ref{sec:exp-patchsqueeze-acc}).

\begin{figure}[bt]
\vspace{-0.2cm}
\centerline{\includegraphics[width=0.9\linewidth]{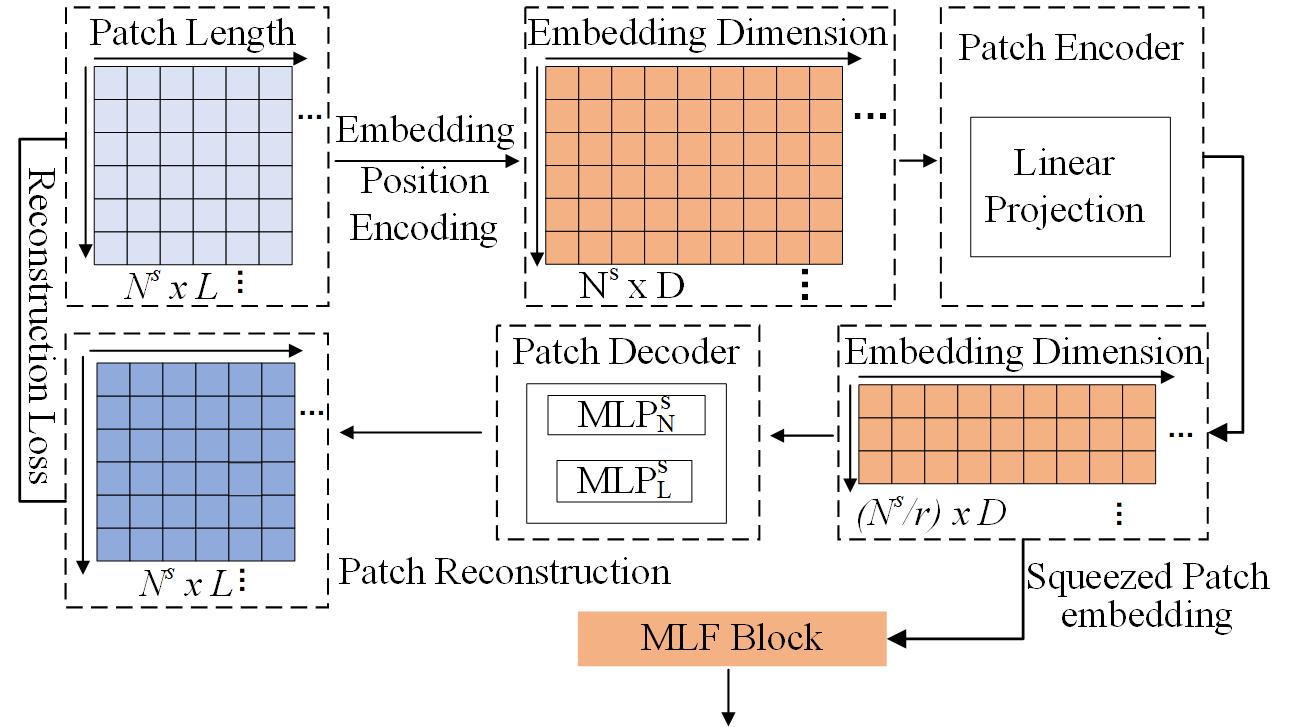} }
\vspace{-0.4cm}
\caption{Illustration of patch squeeze module.}
\vspace{-0.4cm}
\label{fig:patch_squeeze}
\end{figure}

The patch squeeze module, illustrated in Figure~\ref{fig:patch_squeeze}, consists of a lightweight encoder ($PatchEnc$, implemented with one linear layer) and decoder (composed of $MLP_L^s$ and $MLP_N^s$). 
The outputs of the patch squeeze module, denoted as $\mathcal{\hat{X}}_p^{s}$ (used for reconstruction loss) and $\mathcal{\hat{X}}_{d}$ (sent to the transformer encoder), are formulated as follows: 

 \vspace{-0.2cm}
 \begin{equation}
\label{equ:linear_squeeze}
\mathcal{\hat{X}}_d^{s}= PatchEnc(\mathcal{X}_d^{s},r)
\end{equation}
 \begin{equation}
\label{equ:linear_squeeze_concat}
\mathcal{\hat{X}}_d=[\mathcal{\hat{X}}_d^{0},\cdots,\mathcal{\hat{X}}_d^{s}]
\end{equation}
 \begin{equation}
\label{equ:linear_rec}
\mathcal{\hat{X}}_p^{s}= MLP_L^s(MLP_{N}^s(\mathcal{\hat{X}}_d^{s})) 
\end{equation}

\noindent where $r$ represents the squeeze factor, which reduces the original patch number $N^s$ to $\frac{N^s}{r}$. $MLP_N^s$ and $MLP_L^s$ are used to align the dimensions of patch number and patch length accordingly. 
$\mathcal{\hat{X}}_p^{s}$ represents the reconstructed original time series. 
This reconstruction is used to calculate the reconstruction loss, helping the reduced number of patches effectively capture the essential information from the original patches.
$\mathcal{\hat{X}}_{d}$ denotes the concatenation of the squeezed multi-period patch embeddings, which are 
fed into the transformer encoder.

\vspace{-0.1cm}
\subsection{Vanilla Multi-period Attention}
By directly feeding $\mathcal{\hat{X}}_d$ into the transformer encoder, we implement the vanilla multi-period self-attention. Specifically, each head $h = 1, 2, ..., H$ in the multi-head attention processes $\mathcal{\hat{X}}_d$  to generate $H$ distinct query matrices $Q_{h}=(\mathcal{\hat{X}}_d)^{T}W_{h}^Q$, key matrices $K_{h}=(\mathcal{\hat{X}}_d)^{T}W_{h}^K$, and value matrices $V_{h}=(\mathcal{\hat{X}}_d)^{T}W_{h}^V$. 
$T$ denotes matrix transposition. 
A scaled dot-product operation is then applied to compute the attention output $O_{h} \in \mathbb{R}^{D \times N}$:

\vspace{-0.2cm}
 \begin{equation}
\label{equ:Attention}
\vspace{-0.1cm}
O_{h}=Attention(Q_{h},K_{h},V_{h})=\sigma(\frac{Q_{h}(K_{h})^T}{\sqrt{d_k}})V_{h}
\end{equation}

\noindent here, $\sigma$ denotes the Softmax function.
The 
output $z_{e} \in \mathbb{R}^{D \times N}$ at the $e$-th encoder layer is computed using BatchNorm and a feed-forward network with residual connections~\cite{vaswani2017attention,nie2022time_patchformer}.  

\vspace{-0.2cm}
\subsection{Inter-period Redundancy Filtering-based Multi-period Self-attention}
The information redundancy between different periods can cause self-attention to excessively focus on repetitive segments due to higher correlations between similar parts. 
This can prevent self-attention from capturing all patterns distributed across short and medium-/long-term periods, thereby impairing TSF performance. 

To mitigate this issue, we propose an Inter-period Redundancy Filtering (IRF) block, illustrated in Figure~\ref{fig:frame_work}(b). 
IRF identifies redundant information in longer periods based on shorter ones, and then removes it in the embedding space. 
Consequently, IRF helps better self-attention modeling, allowing important information from all periods to be effectively utilized, thereby improving prediction accuracy. 
For implementation, IRF starts by splitting the concatenated  period representations from $z_e$:
 \begin{equation}
\label{equ:final_forecast_z_split}
z^{1}_e,z^{2}_e,\cdots,z^{s}_e,\cdots,z^{S}_e=split(z_e)
\end{equation}
Each $z^{s}_e\in \mathbb R^{D \times N^s}$ denotes the representation of the $s$-th period. 
Then, we devise a Single Period Processing (SPP, as shown in Figure~\ref{fig:frame_work}(b)) module for each period representation $z^{s}_{e}$. 
SPP has two branches of linear layers.
The forecast branch generates prediction of the current period, while the redundancy estimation branch estimates the redundancy to be removed from the longer period representation $z^{s'}_{e}$ ($s < s' \leq S$).
SPP outputs the forecasts $\mathcal{\hat{X}}^{s}_{f_e}$ and the estimated redundancy $\epsilon^{s}_e$ for the $s$-th period at $e$-th block.

 \begin{equation}
\label{equ:final_forecast_FB}
\mathcal{\hat{X}}^{s}_{f_{e}},\epsilon^{s}_e=SPP(z^{s}_e)
\end{equation}
To reduce redundancy, we subtract the estimated redundancy of all shorter periods from the current period embedding $z^{s}_e$.
The period representation after redundancy filtering is denoted as $\hat{z}^{s}_e$:

 \begin{equation}
\label{equ:noise_filter}
\hat{z}^{s}_e=z^{s}_e-\sum_{j=0}^{s-1} \frac{\epsilon^{j}_e}{\sqrt{d_k}} 
\end{equation}
$\sqrt{d_k}$ serves as a scaling factor designed to prevent numerical overflow and stabilize gradients.
$\hat{z}^{s}_e$ are then output to the subsequent encoder layer.

Lastly, as illustrated in  Figure~\ref{fig:frame_work}(a), stacking the MLF block (one transformer block and one IRF block) 
enables progressive redundancy filtering, thereby improving the accuracy of redundancy estimates and self-attention modeling.
Each MLF block has a forecast head~\cite{DBLP:conf/aistats/LeeXGZT15,huang2017densely}.
We aggregate the forecasts from all blocks by:
\vspace{-0.2cm}
 \begin{equation}
\label{equ:final_forecast_4_enclayers}
\mathcal{\overline{X}}_f^s=\frac{1}{E}\sum_{e=1}^{E} \mathcal{X}^{s}_{f_{e} }
\end{equation}
where $\mathcal{\overline{X}}_f^s$ is the average forecast at the block level for period $s$.

\vspace{-0.1cm}
\subsection{Learnable Weighted-average Integration (LWI) }
After obtaining forecasts $\mathcal{\overline{X}}_f^s$ for each 
period,
we now integrate multiple period forecasts to obtain the final prediction by learned weighted averaging.
As different future positions are best forecasted by different periods, simply adding or averaging these predictions may lead to inaccurate forecasting. 
Hence, we propose a Learnable Weighted-average Integration (LWI) module to effectively integrate multi-period forecasts. 
LWI adaptively assigns the weights for different period forecasts, which increases the contribution of accurate predictions while reducing the impact of inaccurate ones.

To implement LWI, we start by extracting temporal features $\nu$ from the longest-period TS $\mathcal{X}_h^S$ using a lightweight Convolutional Neural Network (CNN), which consists of a convolutional layer ($Conv(\cdot)$), padding function ($Pad(\cdot)$), BatchNorm ($BN(\cdot)$) and Maxpooling ($MaxPool(\cdot)$) layer.
Then we employ two MLPs to derive query and key values, enabling a dot product operation to compute attention scores that indicate the model's effective focus on different periods. 
These attention scores are then passed through a sigmoid activation function $\varsigma(\cdot)$ to generate weights $Att$ for each period. 
Finally, we average the multi-period forecasts based on the learned weights as the final prediction (Eq.~\ref{equ:calibrate_f}). 
The entire process that depicted in right of Figure~\ref{fig:frame_work}(a) can be formulated as follows:

\vspace{-0.1cm}
 \begin{equation}
\label{equ:features_e}
\nu=MaxPool(BN(Conv(Pad(\mathcal{X}_h))))
\vspace{-0.1cm}
\end{equation}
\vspace{-0.2cm}
 \begin{equation}
\label{equ:features_e}
Att=\varsigma\left( \tanh{(\Theta_1\nu+\textbf{b}_1)}\odot\tanh{(\Theta_2\nu+\textbf{b}_2)}\right)
\vspace{-0.1cm}
\end{equation}
\vspace{-0.2cm}
 \begin{equation}
\mathcal{\overline{X}}_f=\frac{1}{S}\sum_{e=1}^{S} \mathcal{X}^{s}_{f}\odot Att^s
\label{equ:calibrate_f}
\end{equation}
where $\Theta$ and $\textbf{b}$ are learnable parameters.
\subsection{System Deployment}
\begin{figure}[bt]
\centerline{\includegraphics[width=\linewidth]{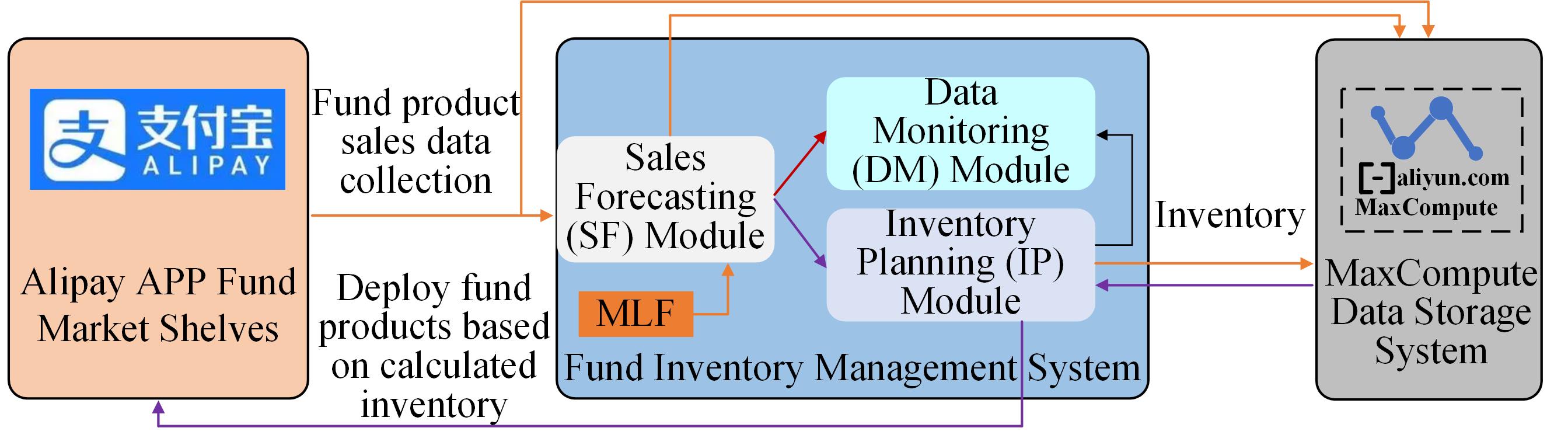} }
\vspace{-0.3cm}
\caption{Alipay's fund management system architecture.
MLF has been deployed in this production environment.
}
\label{fig:deployment_system_3}
\end{figure}

We illustrate the online system architecture in Figure~\ref{fig:deployment_system_3}. 
It consists of three sub-systems: the Fund Inventory Management System (FIMS), Alipay APP Market Shelf, and the MaxCompute database. 
FIMS incorporates modules for Sales Forecasting (SF, with MLF in it), Inventory Planning (IP), and Data Monitoring (DM). 
SF predicts fund sales for the upcoming 5 days based on historical sales data. 
IP uses an inventory planning algorithm to determine product inventory according to the forecasted sales. 
DM monitors the reasonableness of forecasted sales and inventory plans.

The system operates as follows.
The shelf subsystem continuously sends trading data to the FIMS subsystem, which updates the sales records of fund products in the MaxCompute database. 
At the end of each day, FIMS performs three sequential daily batch tasks: 1) retraining the MLF model; 2) using the updated MLF to forecast fund sales for the next 5 days; and 3) planning product inventory with the IP module. 
Subsequently, the inventory of all fund products on the shelf is adjusted accordingly. 
Additionally, both forecasted fund sales and planned inventory are persistently stored in the MaxCompute database for retrospective analysis.

\textbf{Training.}
The final loss function for MLF is:
\vspace{-0.15cm}
\begin{equation}
\label{equ:loss}
    \mathcal{L_{\text{MLF}}}=\mathcal{L_{\text{MSE}}}(\mathcal{\overline{X}}_f,\mathcal{X}_f)+  \frac{1}{S}\sum_{j=1}^{S} 
   \mathcal{L_{\text{MSE}}}(\mathcal{\hat{X}}^{j}_{p},\mathcal{X}^{j}_{p}) 
\end{equation}
\noindent here, the first term refers to the forecasting loss, while the second term represents the reconstruction loss.

\section{Experiments and Results}
\subsection{Experimental Settings}
\subsubsection{Datasets}
\textbf{Fund sales dataset.} 
We collected sales data for different fund products from Ant Fortune, an online wealth management platform on the Alipay APP, consisting of daily user transactions for applying and redeeming funds. 
The dataset spans from January 2015 to January 2023 and is split into training, validation, and testing sets in a 7:1:2 ratio. 
For model training and evaluation, we merged the training, validation, and test sets across all fund datasets.

\textbf{Public datasets.} These datasets, as shown in Table~\ref{tab:dataset_stat}, cover a range of time steps and variables and have been widely employed in the literature for multivariate forecasting tasks~\cite{nie2022time_patchformer,wu2021autoformer,zhou2022fedformer}.


 \begin{table} [bt] 
    \setlength{\tabcolsep}{1.5pt}
    \centering
    \vspace{-0.2cm}
    \caption{Statistics of used popular public datasets.}
    \vspace{-0.3cm}
    \label{tab:dataset_stat}
    { \small
    \begin{tabular}{c|c|c|c|c|c|c}
        \hline
        \multirow{1}{*}{\shortstack{Datasets}}  &  \multicolumn{1}{c|}{ETTh1/h2} &  \multicolumn{1}{c|}{ETTm1/m2}  & \multicolumn{1}{c|}{Weather}& \multicolumn{1}{c|}{Exchange}& \multicolumn{1}{c|}{Illness}& \multicolumn{1}{c}{Electricity}\\
         \midrule[0.5pt]
         \multirow{1}{*}{Features}  &7 &7 &21 &8&7&321\\
        \midrule[0.5pt]
         \multirow{1}{*}{Timesteps}  &17420 &69680 &52696&7588&966&26304\\
        \midrule[0.5pt]
    \end{tabular}
    } 
\vspace{-0.5cm}
\end{table}

\begin{table*}[h] 
    \setlength{\tabcolsep}{2.75pt}
    \centering
    \vspace{-0.05cm}
    \caption{Comparing MLF with popular benchmarks in short-term multivariate forecasting.  ``-'' denotes that the prediction length is too short to use Scaleformer. The best results are in bold and the second best are underlined. }
    \vspace{-0.15cm}
    \label{tab:metric_fund}
    { \small
    \begin{tabular}{c|c|cc|cc|cc|cc|cc|cc|cc|cc}
        \hline
        \multirow{2}{*}{\shortstack{}} & &  \multicolumn{2}{c|}{MLF} &  \multicolumn{2}{c|}{PatchTST}&  \multicolumn{2}{c|}{Patch-Ensemble} & \multicolumn{2}{c|}{Patch-Concat}& \multicolumn{2}{c|}{NHits} &  \multicolumn{2}{c|}{FiLM}&  \multicolumn{2}{c|}{Scaleformer} & \multicolumn{2}{c}{Pathformer}\\
         & & MSE & WMA. & MSE & WMA. & MSE & WMA.  & MSE & WMA.& MSE & WMA.& MSE & WMA.& MSE & WMA.& MSE & WMA. \\ 
         \midrule[0.5pt]
         \multirow{8}{*}{Fund} &1 &\textbf{33.85} &\textbf{75.84}&36.68 &81.05&40.30 &86.68 &39.68 &85.87&36.53 &\underline{80.29}&46.12 &96.10&-&-&\underline{35.54} &80.75\\
         
         &&\textbf{$\pm$0.033} &\textbf{$\pm$0.093}&$\pm$0.068 &$\pm$0.288&$\pm$0.235 &$\pm$0.431&$\pm$0.328 &$\pm$0.266&$\pm$0.057 &$\pm$0.108&$\pm$0.009 &$\pm$0.009&- &-&$\pm$0.08 &$\pm$0.06\\
       \cline{3-18}
         &5 &\textbf{38.28} &\textbf{80.37}&40.28 &83.09&39.95 &83.24 &41.60 &87.14&40.91 &83.93&42.86 &85.35&40.43 &83.74&\underline{39.90} &\underline{82.58}\\
        &&\textbf{$\pm$0.029} &\textbf{$\pm$0.071}&$\pm$0.055 &$\pm$0.143&$\pm$0.068 &$\pm$0.087&$\pm$0.357 &$\pm$0.643&$\pm$0.024 &$\pm$0.090&$\pm$0.009 &$\pm$0.001&$\pm$0.088 &$\pm$2.47&$\pm$0.02 &$\pm$0.06 \\
        \cline{3-18}
         & 8 &\textbf{41.94} &\textbf{86.06}&44.71 &88.81&44.06 &88.77  &\underline{43.49} &\underline{88.08}&44.81 &89.23&44.57 &89.12&43.97&87.62&44.17 &88.67\\
         &&\textbf{$\pm$0.031} &\textbf{$\pm$0.123}&$\pm$0.085 &$\pm$0.164&$\pm$0.052 &$\pm$0.094 &$\pm$0.289 &$\pm$0.552&$\pm$0.020 &$\pm$0.062&$\pm$0.009 &$\pm$0.001&$\pm$0.062 &$\pm$2.22&$\pm$0.04 &$\pm$0.11 \\

         \cline{3-18}
         & 10 &\textbf{44.42} &\textbf{88.66}&46.18 &90.63&46.09 &91.36 &45.59 &90.46 &46.73 &91.12&46.62 &92.67&45.75 &\underline{89.63}&\underline{45.56} &89.82\\
         &&\textbf{$\pm$0.057} &\textbf{$\pm$0.041}&$\pm$0.063&$\pm$0.160&$\pm$0.049 &$\pm$0.070&$\pm$0.054 &$\pm$0.101&$\pm$0.087 &$\pm$0.097&$\pm$0.004 &$\pm$0.009&$\pm$0.088 &$\pm$2.23&$\pm$0.06 &$\pm$0.13  \\
        
         \midrule[0.5pt]
        & & MSE & MAE & MSE & MAE & MSE & MAE  & MSE & MAE& MSE & MAE& MSE & MAE& MSE & MAE& MSE & MAE \\ 
         \multirow{1}{*}{Electricity}&1 &\textbf{0.0472} &\textbf{0.1340}&0.0510 &0.1410&0.053 &0.144&\underline{0.0492} &\underline{0.1382}&0.0517&0.1430&0.0729&0.1750&-&-&0.0613&0.1590 \\
         \midrule[0.5pt]
         \multirow{1}{*}{ETTh{1}}&1  &\textbf{0.0873} &\textbf{0.1899}&0.0979 &0.2033&0.115 &0.2202 &\underline{0.0958} &\underline{0.1983}&0.113 &0.2268 &0.1327 &0.2395&-&-&0.1067&0.213\\
         \midrule[0.5pt]
          \multirow{1}{*}{ETTm{1}} &1 &\textbf{0.0412} &\textbf{0.1245}&0.0461 &0.13&0.0524 &0.1457 &\underline{0.0450} &\underline{0.1298}&0.0475 &0.139 &0.0998 &0.1918&-&-&0.0470&0.1346 \\
          \midrule[0.5pt]
          \multirow{1}{*}{Illness}&1  &\textbf{0.1493} &\textbf{0.2188}&\underline{0.1782} &\underline{0.2340}&0.2918 &0.2934 &0.2476 &0.2409&0.2515 &0.2833 &0.3104 &0.3306&-&-&0.269&0.279\\
          \midrule[0.5pt]
        \multirow{1}{*}{Exchange} &1 &\textbf{0.0029} &\textbf{0.0267}&0.0042 &0.0363&0.0037 &0.030 &0.0047 &0.0325&0.008 &0.0559 &0.0055 &0.0445&-&-&\underline{0.0035}&\underline{0.0293}\\
        \midrule[0.5pt]
    \end{tabular}
    } 
\vspace{-0.1cm}
\end{table*}

\begin{figure*}[h]
\centerline{\includegraphics[width=1.0\linewidth]{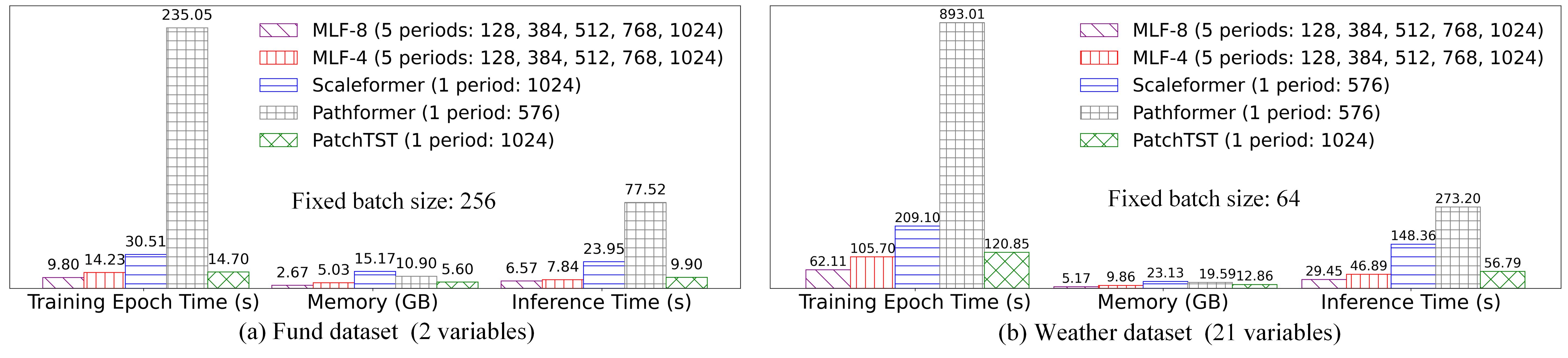} }
\vspace{-0.3cm}
\caption{
Efficiency analysis for best-performing TSF models was conducted on Ettm1, Weather, and Fund datasets.}
\vspace{-0.1cm}
\label{fig:speed_memory_0531}
\end{figure*}

\vspace{-0.1cm}
\subsubsection{Evaluation Metrics} 
We evaluate multivariate TSF tasks using three metrics: Mean Squared Error (MSE), Mean Absolute Error (MAE), and Weighted Mean Absolute Percentage Error (WMAPE or WMA. for short). 
WMAPE is particularly relevant for the fund dataset on the Alipay APP, as it assesses the accuracy of predictions on larger values. 
This metric aligns with the business goal of accurately forecasting larger transactions in fund products. 
We present the sum of WMAPE for the two variables in the fund dataset.

\begin{table*}[h]
    \centering
    \vspace{-0.05cm}
    \caption{Comparing MLF with popular benchmarks in long-term multivariate forecasting. The best results are indicated in bold, and the second-best results are underlined.
    }
    \vspace{-0.2cm}
    \label{tab:metric_public_long_term}
    { \small
    \begin{tabular}{c|c|cc|cc|cc|cc||cc|cc|cc|cc}
        \hline
        \multirow{2}{*}{\shortstack{}} & &  \multicolumn{2}{c|}{MLF} &  \multicolumn{2}{c|}{PatchTST} & \multicolumn{2}{c|}{Patch-Ensemble} & \multicolumn{2}{c||}{Patch-Concat}&  \multicolumn{2}{c|}{NHits} &  \multicolumn{2}{c|}{FiLM} & \multicolumn{2}{c|}{Scaleformer} & \multicolumn{2}{c}{Pathformer}\\
         & & MSE & MAE & MSE & MAE & MSE & MAE  & MSE & MAE& MSE & MAE& MSE & MAE& MSE & MAE& MSE & MAE \\ 
         \midrule[0.5pt]
         \multirow{4}{*}{\rotatebox[origin=c]{90}{ETTh1}} 
         &96 &\textbf{0.363} &\textbf{0.392}&0.370 &0.400&0.370 &0.401&\underline{0.369} &\underline{0.400}&0.427 &0.439&0.371 &0.394&0.379 &0.409&0.393 &0.406\\
         &192 &\textbf{0.399} &\textbf{0.416}&0.413 &0.429&\underline{0.402} &\underline{0.422}&0.406 &0.427&0.472 &0.473&0.414 &0.423&0.411 &0.43&0.421 &0.436\\
         & 336 &\textbf{0.416} &\textbf{0.423}&\underline{0.422} &0.440&0.424 &\underline{0.432}&0.427 &0.433&0.525 &0.498&0.442 &0.445&0.43 &0.443&0.451 &0.451\\
         & 720 &\textbf{0.438} &\textbf{0.454}&0.447 &0.468&\underline{0.445} &\underline{0.463}&0.445 &0.463 &0.608 &0.565&0.465 &0.472&0.446 &0.465&0.483 &0.470\\
        \midrule[0.5pt]

        \multirow{4}{*}{\rotatebox[origin=c]{90}{ETTh2}} &96 &\textbf{0.267} &\textbf{0.334}&\underline{0.274} &\underline{0.337}&0.276 &0.341&0.277 &0.341&0.314 &0.379&0.284 &0.348&0.275 &0.343&0.285 &0.349\\
         &192 &\textbf{0.327} &\textbf{0.374}&0.341 &0.382&0.337 &0.379&0.336 &\underline{0.379}&0.401 &0.434&0.357 &0.4&0.337 &0.384&\underline{0.331} &0.385\\
         & 336 &\textbf{0.350} &\textbf{0.396}&0.359 &0.405&\underline{0.357} &\underline{0.403}&0.358 &0.403&0.452 &0.469&0.377 &0.417&0.364 &0.414&0.368 &0.409\\
         & 720 &\textbf{0.383} &\textbf{0.423}&0.388 &\underline{0.427}&\underline{0.387} &0.429&0.388 &0.430&0.545 &0.527&0.439 &0.456&0.397 &0.438&0.389 &0.427 \\
        \midrule[0.5pt]
        
        \multirow{4}{*}{\rotatebox[origin=c]{90}{ETTm1}} &96 &\textbf{0.285} &\textbf{0.344}&0.293 &0.346&0.296 &0.347&\underline{0.290} &\underline{0.345}&0.32 &0.367&0.302 &0.345&0.293 &0.347&0.301 &0.352\\
         &192 &\textbf{0.324} &\textbf{0.366}&0.333 &\underline{0.370}&0.332 &0.373&\underline{0.331} &0.372&0.357 &0.392&0.338 &0.368&0.333 &0.371&0.356 &0.383\\
         & 336 &\textbf{0.355} &\textbf{0.384}&\underline{0.360} &\underline{0.392}&0.366 &0.395&0.365 &0.396&0.392 &0.417&0.365 &0.385&0.364 &0.391&0.387 &0.405\\
         & 720 &\textbf{0.401} &\textbf{0.410}&\underline{0.404} &\underline{0.417}&0.415 &0.427&0.417 &0.427&0.442 &0.441&0.42 &0.42&0.42 &0.425&0.416 &0.420 \\
        \midrule[0.5pt]
        
        \multirow{4}{*}{\rotatebox[origin=c]{90}{ETTm2}} &96&\textbf{0.164} &\textbf{0.253}&0.166 &\underline{0.256}&\underline{0.163} &0.257&0.169 &0.263&0.176 &0.255&0.165 &0.256&0.172 &0.255&0.168 &0.258\\
         &192 &\textbf{0.218} &\textbf{0.295}&0.223 &\underline{0.296}&0.226 &0.304&0.227 &0.306&0.245 &0.305&\underline{0.222} &0.296&0.231 &0.298&0.227 &0.296\\
         & 336 &\textbf{0.270} &\textbf{0.330}&0.277 &0.336&0.278 &0.336&0.285 &0.340&0.295 &0.346&0.277 &0.333&0.278 &0.328&\underline{0.273} &\underline{0.331}\\
         & 720 &\textbf{0.344} &\textbf{0.379}&\underline{0.357} &\underline{0.381}&0.360 &0.393&0.360 &0.393&0.401 &0.413&0.371 &0.389&0.361 &0.383&0.366 &0.392 \\
        \midrule[0.5pt]
        
        \multirow{4}{*}{\rotatebox[origin=c]{90}{Weather}} &96 &\textbf{0.145} &\textbf{0.195}&0.148 &0.200&0.149 &0.201&\underline{0.147} &\underline{0.199}&0.158 &0.212&0.199 &0.262&0.152 &0.208&0.155 &0.208\\
         &192 &\textbf{0.191} &\textbf{0.238}&0.194 &0.241&0.196 &0.244&\underline{0.194} &\underline{0.240}&0.202 &0.247&0.228 &0.288&0.197 &0.251&0.196 &0.246\\
         & 336 &\textbf{0.241} &\textbf{0.280}&\underline{0.245} &\underline{0.282}&0.247 &0.284&0.248 &0.284&0.257 &0.3&0.267 &0.323&0.253 &0.296&0.25 &0.286\\
         & 720 &\textbf{0.296} &\textbf{0.328}&\underline{0.309} &\underline{0.330}&0.318 &0.337&0.316 &0.335&0.327 &0.356&0.319 &0.361&0.311 &0.343&0.324 &0.337 \\
        \midrule[0.5pt]

    \end{tabular}
    } 
\vspace{-0.2cm}
\end{table*}
\subsubsection{Baselines} We compare our Multi-period Learning Framework (MLF) against three types of baselines.
\textbf{(i) single-period based models:} This category includes PtachTST~\cite{nie2022time_patchformer}, NHits~\cite{challu2022n_Nhits}, Scaleformer~\cite{shabani2022scaleformer} and PathFormer~\cite{chen2024pathformer}.
Since Scaleformer serves as a general architecture, we combine it with Autoformer~\cite{wu2021autoformer}, NHits~\cite{challu2022n_Nhits}, and PatchTST~\cite{nie2022time_patchformer} to create strong baseline.

\textbf{(ii) Multi-period based models:} This category includes FiLM~\cite{zhou2022film} and two multi-period variants based on PatchTST, called Patch-Concat and Patch-Ensemble. Specifically, for Patch-Concat, we concatenate multi-period data for training PatchTST. 
For Patch-Ensemble, we train multiple instances of PatchTST using multiple periods with different TS lengths for ensemble learning. 
We compare these three baselines to demonstrate the necessity of the structure designed for multi-period forecasting in MLF.

\subsubsection{Implementation Details}
For short-term time series forecasting (TSF) tasks, the short-term to long-term multi-period lengths for MLF are 5, 10, 30, 60, 120, and 150 time steps. 
Prediction lengths vary across $m \in {1, 5, 8, 10}$. 
For long-term TSF tasks,  the used short-term to long-term multi-period lengths are 128,256,512,768, 1024 and 2048 time steps. 
Prediction lengths vary across $m \in {96, 192, 336, 720}$. 
Each period within MLF employs a fixed number of 64 patches, with a squeeze factor $r\in$ {2,4,8} to squeeze the long-sequence input. 
Both Patch-Concat and Patch-Ensemble models utilize the same multi-period configuration as MLF.  
The learning rate and batch size are set to 0.0001 and 128 respectively. 
All methods follow the same data loading parameters (e.g., train/val/test split ratio) as in~\cite{nie2022time_patchformer}. 
Each method is trained for 30 epochs. 
All experiments in this paper are conducted using PyTorch on an NVIDIA GeForce RTX 3090 GPU.

\vspace{-0.2cm}
\subsection{Time Series Forecasting Accuracy}
\subsubsection{Comparison with single-period baselines}
Extensive results show that MLF outperforms advanced single-period baselines, including PatchTST, NHits, Scaleformer, and PathFormer on both short-term (Table~\ref{tab:metric_fund}) and long-term tasks (Table~\ref{tab:metric_public_long_term}). 
These baselines are well-designed for single-period time series and doesn't utilize multi-period information. 
However, we have observed that inputs of different lengths have a significant impact on prediction accuracy. However, these single-period-based methods can only process one input at a time, which prevents them from achieving better prediction accuracy.
In contrast, MLF not only utilizes multi-period inputs, but also incorporates new designs to maximize its TSF potential, achieving better accuracy.
\vspace{-0.1cm}
\subsubsection{Comparison with multi-period methods} 
Extensive results also show that MLF outperforms advanced multi-period methods FiLM, Patch-Concat and Patch-Ensemble, as shown Table~\ref{tab:metric_fund} and Table~\ref{tab:metric_public_long_term}.
This highlights that merely linearly integrating multi-period outputs (FiLM), concatenating multi-period inputs for training PatchTST (Patch-Concat) and using multiple instances of PatchTST for ensemble learning (Patch-Ensemble) fails to fulfill the forecasting potential of multi-period data. 
In contrast, MLF is customized to address the characteristics of multi-period data, such as inter-period redundancy and varying sequence lengths. 
New designs in MLF enable it to effectively utilize information across all periods, thereby achieving superior accuracy in TSF tasks.

\subsection{Time Series Forecasting Efficiency}
\label{sec:exp-efficiency}

In MLF, we use squeeze factors of 8 (MLF-8) and 4 (MLF-4). 
For a fair comparison, we maintained consistency in shared hyper-parameters (such as hidden size and attention heads in the transformer) and settings (like batch size). 

We select the best-performing models for an efficiency comparison. 
Results are depicted in Figure~\ref{fig:speed_memory_0531}. 
Due to GPU memory constraints (24GB),  baselines like Pathformer can handle a maximum sequence length of 576 across all datasets, while Scaleformer manages 576 for Weather datasets and 1024 for ETTm1 or Fund datasets. 
Despite MLF's significantly longer actual input sequence length (2816 compared to 576 or 1024 for PathFormer and Scaleformer), it achieves better efficiency. 
For instance, on the Fund dataset (Figure~\ref{fig:speed_memory_0531}(a)), MLF is 3.6 times faster than Scaleformer and 11.8 times faster than PathFormer in terms of inference time. 
MLF's efficiency advantage is even bigger 
with larger datasets, as shown in the results of the Weather dataset (Figure~\ref{fig:speed_memory_0531}(b)), while maintaining satisfactory accuracy (Table~\ref{tab:metric_public_squeeze_0203}).

\vspace{-0.2cm}
\subsection{Deployment Results}
MLF has been deployed in the FIMS of Alipay since late February 2024. 
We compared the WMAPE and Gross Merchandise Volume (GMV) of all fund products between MLF and the previously deployed model PatchTST. 
Table~\ref{tab:Deployment_results} shows the improvement ratio of WMAPE ($IR_{WMAPE}=\frac{PatchTST_{WMAPE}-MLF_{WMAPE}}{PatchTST_{WMAPE}}$) and the improvement ratio of GMV ($IR_{GMV}=\frac{MLF_{GMV}-PatchTST_{GMV}}{PatchTST_{GMV}}$) over five consecutive weeks. 
The improvement ratios of WMAPE and GMV over advanced PatchTST demonstrate that multi-period based MLF is effective in the fund sales forecasting scenario.

  \begin{table}[bt]
    \setlength{\tabcolsep}{3pt}
    \centering
    \caption{Deployment results over consecutive five weeks.}
    \vspace{-0.1cm}
    \label{tab:Deployment_results}
    { \small
    \begin{tabular}{c|c|c|c|c|c|c}
        \hline
        \multirow{1}{*}{\shortstack{Year (2024)}}  &  \multicolumn{1}{c|}{Week 1} &  \multicolumn{1}{c|}{Week 2}  & \multicolumn{1}{c|}{Week 3}& \multicolumn{1}{c|}{Week 4}& \multicolumn{1}{c|}{Week 5}& \multicolumn{1}{c}{Mean}\\
         \midrule[0.5pt]
         \multirow{1}{*}{$IR_{WMAPE}$ (\%)}  &6.84 &4.25 &5.21 &4.92&4.62&5.16\\
        \midrule[0.5pt]
         \multirow{1}{*}{$IR_{GMV}$ (\%)}  &1.14 &0.
         58&0.72&0.85&0.95&0.84\\
        \midrule[0.5pt]
    \end{tabular}
    } 
\vspace{-0.3cm}
\end{table}

\vspace{-0.1cm}
\subsection{Ablation Study}
\subsubsection{Ablation of each component (Table~\ref{tab:ablation})}
\textbf{(1) Effectiveness of Inter-period Redundancy Filtering (IRF).} ``w/o IRF'' denotes we remove redundancy subtract operation (without using Eq.~\ref{equ:noise_filter}) in IRF block. 
When removing IRF, we observe raised MSE. 
Moreover, compared to "w/o MA", "w/o (MA+IRF)" also shows further raised MSE. 
The raised MSE demonstrates that filtering redundancy between periods allows self-attention to avoid overly focusing on repetitive parts. 
This enables the utilization of information from all periods, resulting in better TSF accuracy. 

Figures~\ref{fig:attn_ASS}(a)-(d) further illustrate this. 
Specifically, we obtained self-attention heatmaps by averaging the attention score matrices of all test samples on the Fund dataset. 
Each index on the horizontal axis represents a patch, with every consecutive three patches representing a period. 
The heatmap without the IRF operation shows an overemphasis on the overlapping parts among the multi-period inputs (vertical highlight areas in Figures~\ref{fig:attn_ASS}(b) and (d)), while the heatmap with the IRF operation demonstrates that most regions (non-repetitive parts among periods) receive effective attention.

\begin{table}[bt]
    \centering
    \small
    \caption{Ablation study on various datasets. 
    }
    \vspace{-0.2cm}
    \label{tab:ablation}
    { \small
    \begin{tabular}{c|c|c|c|c|c}
        \hline
        \multirow{2}{*}{\shortstack{Methods/\\Datasets}} &   \multicolumn{1}{c|}{Illness} &   \multicolumn{1}{c|}{Electricity}&   \multicolumn{1}{c|}{ETTh{1}} &   \multicolumn{1}{c|}{Exchange}&   \multicolumn{1}{c}{Fund}\\ 
         & MSE & MSE  & MSE  & MSE & WMA.\\ 
        \midrule[0.5pt]
        \multirow{1}{*}{MLF} &\textbf{0.149} &\textbf{0.0472} &\textbf{0.087} &\textbf{0.0030} &\textbf{75.84}  \\

        \midrule[0.5pt]
        \multirow{1}{*}{w/o IRF} &0.163 &0.0500 &0.091 &0.0033  &78.56 \\

        \midrule[0.5pt]
        \multirow{1}{*}{\shortstack{w/o (LWI)}} &0.155 &0.0491 &0.088 &0.0032  &77.32 \\

        \midrule[0.5pt]
        \multirow{1}{*}{w/o MA} &0.236 &0.0539 &0.114 &0.004  &78.23 \\

        \midrule[0.5pt]
        \multirow{1}{*}{\shortstack{w/o (MA+IRF)}} &0.261 &0.0559 &0.130 &0.004  &79.85 \\

        \hline
      \end{tabular}
      } 
\vspace{-0.1cm}
\end{table}

\noindent \textbf{(2) Effectiveness of MA and LWI.} 
The raised MSE observed in the "w/o LWI"  and "w/o MA" highlights the importance of the Learnable Weighted-average Integration (LWI) and Multi-period self-Attention (MA) modules.
Based on the learned adaptive period weights, LWI increases the contribution of accurate predictions while reducing the impact of inaccurate ones, thereby improving overall forecasting accuracy. 
MA effectively captures multi-period dependencies. 
Together, these modules are crucial for realizing the full potential of multi-period based time series forecasting. 

\subsubsection{Effectiveness of Multi-period self-Adaptive Patching (MAP)}
\begin{figure}[h]
\centerline{\includegraphics[width=0.6\linewidth]{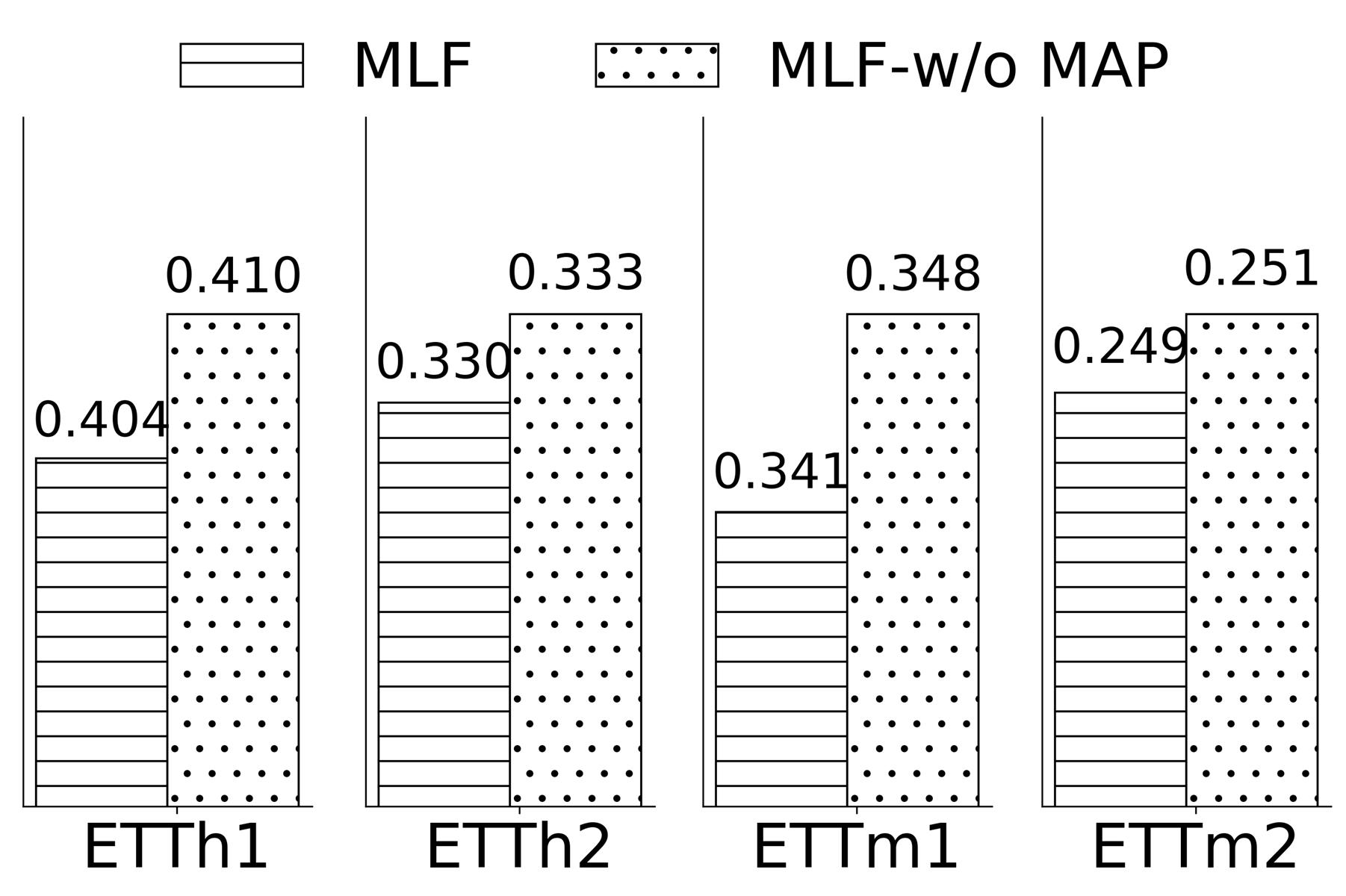} }
\vspace{-0.2cm}
\caption{
Ablation study of Multi-period self-Adaptive Patching (MAP) module. We report the average MSE values for predicting future 96, 192, 336, and 720 time steps.}
\vspace{-0.2cm}
\label{fig:Ablation_MAP}
\end{figure}

In the standard multi-period patching, the shorter periods have fewer patches while the longer one is the opposite. 
The results in Figure~\ref{fig:Ablation_MAP} show the raised MSE when MAP is removed (w/o MAP), indicating that shorter periods may be unfairly treated in standard multi-period patching due to the insufficient number of patches. 
MAP ensures each period $\mathcal{X}_h^s$ has an equal number of patches, allowing the model to equally pay attention to the time semantics of all periods and improving TSF performance.

\subsubsection{Effectiveness of patch squeeze module}
\label{sec:exp-patchsqueeze-acc}
\begin{figure}[h]
\setlength{\abovecaptionskip}{0.1cm}
\centerline{\includegraphics[width=\linewidth]{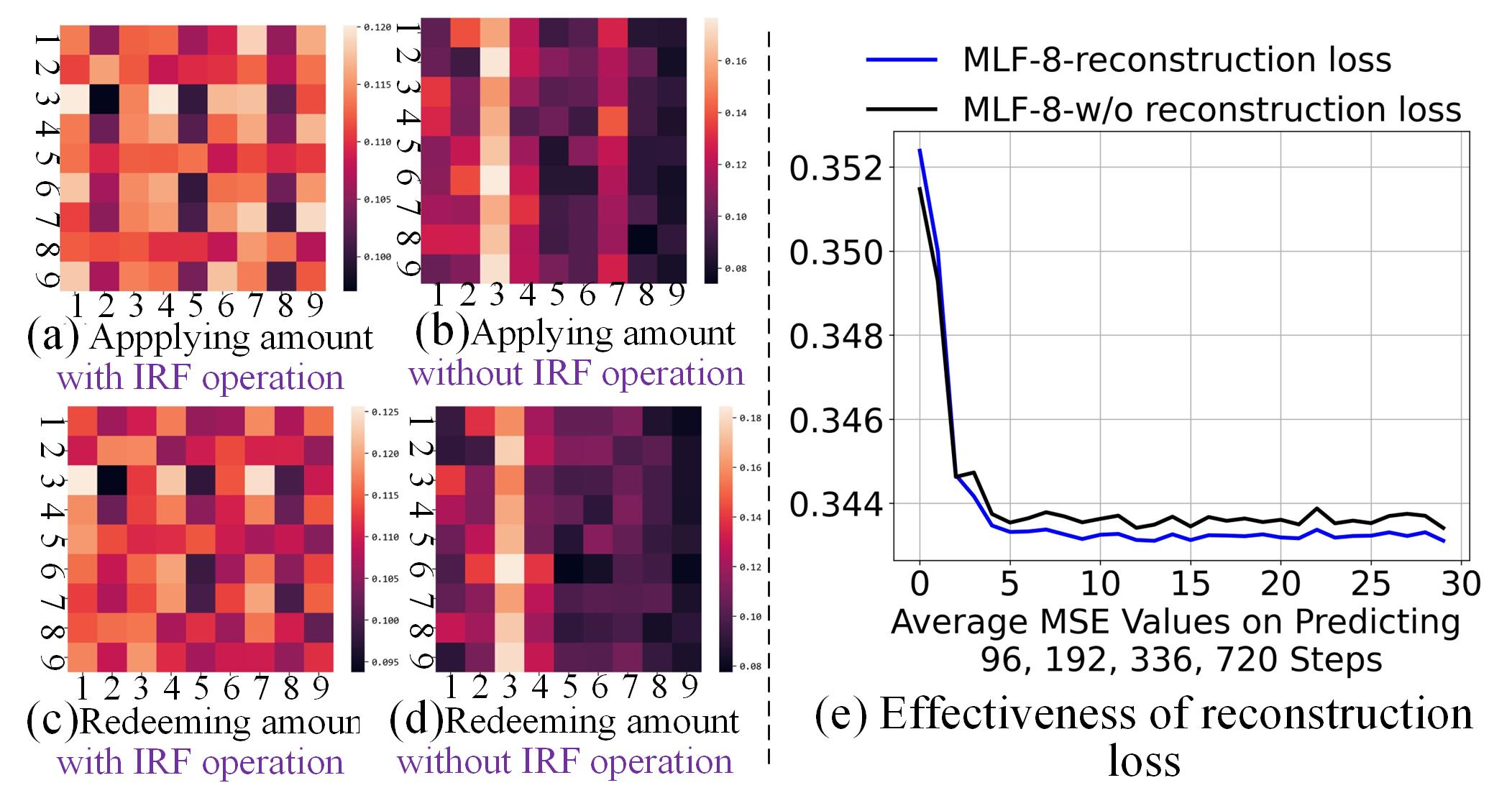} }
\caption{Extra ablation study: (a)-(d) Self-attention visualization with and without IRF. (e) Ablation of the reconstruction loss in the patch squeeze module on the ETTm1 dataset.}
\vspace{-0.1cm}
\label{fig:attn_ASS}
\end{figure}

As shown in Table~\ref{tab:metric_public_squeeze_0203}, MLF achieves satisfactory performance with squeeze factors of 2, 4, and 8. 
This demonstrates that the patch squeeze module in MLF can significantly reduce time complexity while maintaining good accuracy. 
Additionally, the reconstruction loss in the patch squeeze module is effective, as Figure~\ref{fig:attn_ASS}(d) shows raised MSE when this term is removed ("w/o reconstruction loss").

\begin{table}[bt]
    \setlength{\tabcolsep}{3.75pt}
    \centering
    \caption{Effectiveness of the patch squeeze module: MLF's accuracy with squeeze factors 2 (MLF-2), 4 (MLF-4), and 8 (MLF-8). '-' indicates out of memory.}
    \vspace{-0.1cm}
    \label{tab:metric_public_squeeze_0203}
    { \small
    \begin{tabular}{c|c|cc|cc|cc|cc} 
        \hline
        \multirow{2}{*}{\shortstack{}} & &  \multicolumn{2}{c|}{PatchTST} &  \multicolumn{2}{c|}{MLF-2}&  \multicolumn{2}{c|}{MLF-4}& \multicolumn{2}{c}{MLF-8}\\
         & & MSE & MAE  & MSE & MAE  & MSE & MAE  & MSE & MAE\\ 
         \midrule[0.5pt]
 
        \multirow{4}{*}{\rotatebox[origin=c]{90}{ETTm1}}&96 &0.293 &0.346 &\textbf{0.283} &\textbf{0.342} &\underline{0.285} &\textbf{0.344} &0.287 &\underline{0.345}\\
        
         &192 &0.333 &0.370 &\textbf{0.322} &\textbf{0.366}&\underline{0.324} &\textbf{0.366} &0.326 &\textbf{0.366}\\
         
         & 336 &0.360 &0.392 &0.357 &0.387 &\underline{0.356} &\underline{0.384} &\textbf{0.355} &\textbf{0.384}\\
         
         & 720 &0.404 &0.417 &0.406 &0.413 &\textbf{0.400} &\underline{0.410} &\underline{0.401} &\textbf{0.410} \\
        \midrule[0.5pt]

        \multirow{4}{*}{\rotatebox[origin=c]{90}{Weather}} &96 &0.148 &0.200 &- &- &\underline{0.146} &\underline{0.196} &\textbf{0.145} &\textbf{0.195}\\
        
         &192 &0.194 &0.241 &- &- &\textbf{0.191}&\textbf{0.238} &\underline{0.191} &\underline{0.239}\\
         
         & 336 &0.245 &\underline{0.282} &- &-&\underline{0.244} &0.283 &\textbf{0.241} &\textbf{0.280}\\
         
         & 720 &0.309 &0.330 &- &- &\underline{0.304} &\underline{0.333} &\textbf{0.296} &\textbf{0.328}\\
        \midrule[0.5pt]
    \end{tabular}
    } 
\vspace{-0.1cm}
\end{table}

\section{Conclusion}
This paper introduces Multi-period Learning Framework MLF, which incorporates multiple inputs with varying lengths to achieve better accuracy and reduces the costs of selecting input windows during training. MLF also demonstrates that using a simple encoder to squeeze the input time series (dimensionality reduction) can significantly improve model efficiency, reduce memory overhead, and achieve comparable or even better forecasting accuracy. 
MLF's other designs, including MAP, IRF, and LWI also help better address multi-period characteristics and enhance TSF performance. 
Nevertheless, there is still considerable room for exploration in model architectures that use historical inputs of different lengths simultaneously for prediction.
Future research could explore novel designs aimed at better addressing the challenges caused by multi-period characteristics for multi-period forecasting, thereby further releasing the predictive potential of multi-period inputs. To facilitate research, we have open-sourced the code.

\section{Acknowledgements}
This work is supported by the Ministry of Science and Technology of China, National Key Research and Development Program (No. 2021YFB3300503).

\bibliographystyle{ACM-Reference-Format}
\balance
\bibliography{ref}

\appendix
\section{Appendix}

\subsection{More details about fund datasets}
\label{appendix:datasets}
\begin{figure}[bt]
\centerline{\includegraphics[width=\linewidth,height=40mm]{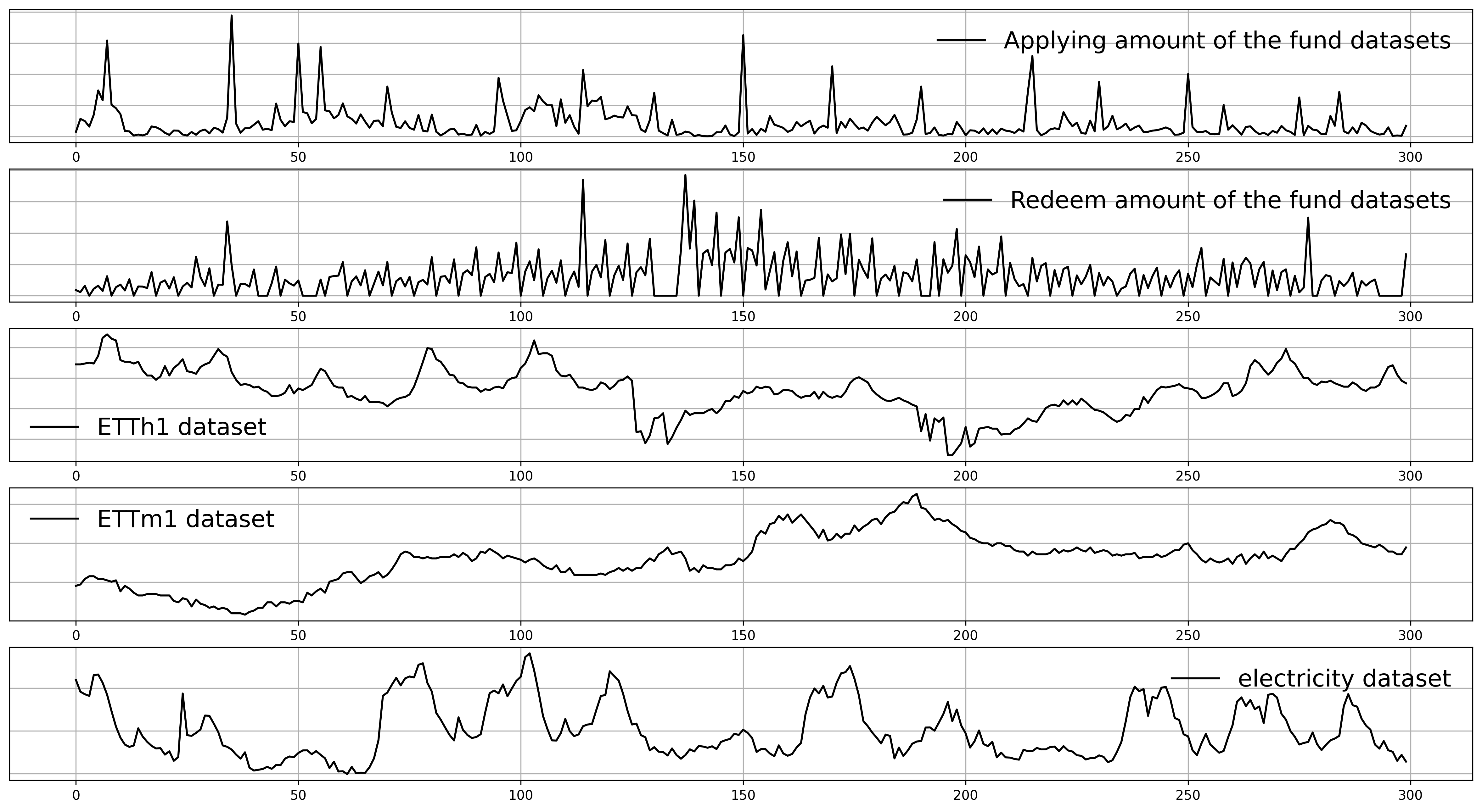} }
\caption{Time series visualization of Fund dataset (first two lines) and other public datasets.}
\label{fig:series_comp_appen}
\end{figure}

On the one hand, from the time series visualization in Figure~\ref{fig:series_comp_appen}, we can intuitively observe pattern differences between the Fund dataset and the public datasets. Influenced by long-term market conditions, short-term policy, and public opinion, the time series of fund sales exhibit steeper ascending and descending trends as well as more frequent fluctuations, containing complex semantics.

On the other hand, inspired by~\cite{wang2006characteristic}, in Table~\ref{tab:data_comp_ana}, we quantify the trend, periodicity, and kurtosis of the Fund dataset and other popular TSF public datasets. Fund dataset exhibits lower trend and periodicity along with higher kurtosis, indicating more complex TS~\cite{wang2006characteristic} and bringing more challenges for TSF tasks. 

By introducing the Fund datasets, we intend to expand the public dataset repository for more comprehensive evaluations of TSF algorithms. The features of each fund product are as follows:

\begin{enumerate}
\item ``product\_pid'' denotes the fund ID of the different fund products. 
\item {``is\_summarydate''} denotes whether the transaction date of the fund product is a \textit{summary date} (fund products do not trade on holidays and weekends, and the trading volume during these periods is aggregated to the next non-holiday or non-weekend, called \textit{summary date}). 

\item {``apply\_amt''} represents the applying transaction amount of the current fund product. 
\item {``redeem\_amt''} represents the redemption transaction amount of the current fund product.  
\item {``during\_days''} indicates the holding period of the current fund product (the number of days to hold the fund product before it can be traded). 
\item {``is\_trad''} indicates whether the current day is a trading day. 
\item {``is\_weekend\_delay''} indicates whether it is a weekend before a trading day. 
\item {``holiday\_num''} indicates how many statutory holidays occur before the trading day.
\end{enumerate}

\begin{table}[bt]
    \setlength{\tabcolsep}{4pt}
    \centering
    \caption{Average trends, periodicity, and kurtosis comparison.}
    \label{tab:data_comp_ana}
    {\small
    \begin{tabular}{c|c|c|c|c|c|c}
        \hline
        \multirow{1}{*}{\shortstack{Datasets}}  &  \multicolumn{1}{c|}{Fund} &  \multicolumn{1}{c|}{Electricity}&  \multicolumn{1}{c|}{ETTh{1}}  & \multicolumn{1}{c|}{ETTm{1}}& \multicolumn{1}{c|}{Exchange}& \multicolumn{1}{c}{Illness}\\
         \midrule[0.5pt]
         \multirow{1}{*}{Trend}  &0.74 &0.84&0.86 &0.86 &1.0&0.71\\
        \midrule[0.5pt]
         \multirow{1}{*}{Periodicity}  &0.62 &0.92&0.72 &0.74&0.31&0.81\\
        \midrule[0.5pt]
        \multirow{1}{*}{Kurtosis}  &29.21 &0.39 &1.37 &1.37 &-0.53&5.56\\
        \midrule[0.5pt]
    \end{tabular}}
\vspace{-0.2cm}
\end{table}

\subsection{More details about public datasets}
The following public datasets are extensively used for TSF algorithm evaluation, covering four practical applications: energy, economics, weather, and disease:

\begin{enumerate}
\item The Electricity dataset collects the electricity consumption (kWh) every 15 minutes of 321 clients from 2012 to 2014.
\item The ETT dataset comprises two sub-datasets, ETT1 and ETT2, collected from two separate counties. Each sub-dataset offers two versions with varying sampling resolutions (15 minutes and 1 hour). ETT dataset includes multiple time series of electrical loads and a single time sequence of oil temperature.
\item The Weather dataset contains 21 meteorological indicators, such as air temperature, humidity, etc, recorded every 10 minutes for the entirety of 2020. 
\item Exchange dataset contains real-time exchange rates for eight different countries.
\item The Illness dataset contains data on patients with influenza-like illnesses in the United States.
\end{enumerate}

\textbf{We have made the Fund and public datasets available at the link}\footnote{\url{https://drive.google.com/drive/folders/1IecxNQqH6hYEgaZT70t273BFHV-WeIw1?usp=sharing}}. Our code link\footnote{\url{https://github.com/Meteor-Stars/MLF}} provides the all data processing scripts.

\begin{table}[h!]
    \setlength{\tabcolsep}{1.75pt}
    \centering
    \caption{More comparison results of short-term TSF task on the collected Fund dataset. }
    \label{tab:metric_fund_appendix}
    {\footnotesize
    \begin{tabular}{c|c|cc|cc|cc|cc|cc}
        \hline
        \multirow{2}{*}{\shortstack{}} & &  \multicolumn{2}{c|}{MLF} &  \multicolumn{2}{c|}{Autoformer}&  \multicolumn{2}{c|}{FEDformer}&  \multicolumn{2}{c|}{LSTM}&  \multicolumn{2}{c}{RNN}\\
         &   & MSE & WMA.& MSE & WMA.& MSE & WMA.& MSE & WMA.& MSE & WMA. \\ 
         \midrule[0.5pt]
         \multirow{8}{*}{\rotatebox[origin=c]{90}{Fund}} &1 &\textbf{33.85} &\textbf{75.84}&36.5 &82.17&36.85 &80.31&51.41 &85.91&\underline{36.2} &\underline{79.82}\\
         
         &&\textbf{$\pm$0.033} &\textbf{$\pm$0.093}&$\pm$0.027 &$\pm$0.007&$\pm$0.002 &$\pm$0.003&$\pm$0.18 &$\pm$0.18&$\pm$0.08 &$\pm$0.06\\
       
         &5 &\textbf{38.28} &\textbf{80.37}&45.45 &93.79&47.92 &97.4&62.98 &101.88&\underline{40.42} &\underline{83.52}\\
        &&\textbf{$\pm$0.029} &\textbf{$\pm$0.071}&$\pm$0.038 &$\pm$0.004&$\pm$0.038 &$\pm$0.003&$\pm$0.18 &$\pm$0.17&$\pm$0.02 &$\pm$0.06 \\
         & 8 &\textbf{41.94} &\textbf{86.06}&51.05 &99.9&45.2 &90.3&69.94 &109.04&44.69 &89.74\\
         &&\textbf{$\pm$0.031} &\textbf{$\pm$0.123}&$\pm$0.144 &$\pm$0.007&$\pm$0.001 &$\pm$0.001&$\pm$0.1 &$\pm$0.03&$\pm$0.04 &$\pm$0.11 \\

         & 10 &\textbf{44.42} &\textbf{88.66}&52.72 &100.35&47.1 &91.92&71.68 &109.78&46.67 &91.55 \\
         &&\textbf{$\pm$0.057} &\textbf{$\pm$0.041}&$\pm$0.104 &$\pm$0.003&$\pm$0.037 &$\pm$0.001&$\pm$0.11 &$\pm$0.08&$\pm$0.06 &$\pm$0.13 \\
        \midrule[0.5pt]

        \multirow{2}{*}{\shortstack{}} &  & \multicolumn{2}{c|}{DLinear}& \multicolumn{2}{c|}{GRU}& \multicolumn{2}{c|}{XGBoost}& \multicolumn{2}{c}{Prophet}& \multicolumn{2}{c}{ARIMA} \\
         &     & MSE & WMA.& MSE & WMA.& MSE & WMA.& MSE & WMA.& MSE & WMA. \\ 
         \midrule[0.5pt]
         \multirow{8}{*}{\rotatebox[origin=c]{90}{Fund}} &1 &97.23 &158.72&51.57 &85.35 &49.29 &93.07&66.5 &128.18&46.64 &103.71\\
         &&$\pm$0.009 &$\pm$0.002 &$\pm$0.03 &$\pm$0.01&-&-&-&-&-&-\\
       
         &5 &99.50 &159.95&59.21 &93.95&53.55 &95.86&93.75 &154.62&49.56 &105.51 \\
        &&$\pm$0.004 &$\pm$0.003&$\pm$0.15 &$\pm$0.23&-&-&-&-&-&- \\

         & 8 &95.37 &148.25 &66.54 &102.56 &57.23 &101.97&136.97 &180.16&53.11 &108.6\\
         &&$\pm$0.005 &$\pm$0.004&$\pm$0.08 &$\pm$0.11 &-&-&-&-&-&- \\

         & 10 &90.70 &142.89&69.0 &104.86 &58.53 &101.7&171.59 &194.13&54.35 &108.91\\
         &&$\pm$0.005 &$\pm$0.004&$\pm$0.07 &$\pm$0.11&-&-&-&-&-&-  \\

        \midrule[0.5pt]
    \end{tabular}
    }
\end{table}

 \begin{table*}[h!]
    \setlength{\tabcolsep}{4pt}
    \centering
    \caption{Extra result of short-term TSF task on the public datasets.}
    \label{tab:metric_public_appendix}
    {\small
    \begin{tabular}{c|cc|cc|cc|cc|cc|cc|cc}
        \hline
  
        \multirow{2}{*}{}  &  \multicolumn{2}{c|}{MLF} & \multicolumn{2}{c|}{Autoformer}& \multicolumn{2}{c|}{FEDformer}& \multicolumn{2}{c}{LSTM} & \multicolumn{2}{c|}{DLinear}& \multicolumn{2}{c|}{RNN}& \multicolumn{2}{c}{GRU}\\
          & MSE & MAE & MSE & MAE & MSE & MAE  & MSE & MAE& MSE & MAE& MSE & MAE& MSE & MAE \\ 
         \midrule[0.5pt]

        \multirow{1}{*}{ETTh{1}}  &\textbf{0.0873} &\textbf{0.1899}&0.2213 &0.3429&0.2318 &0.3495&0.1318 &0.2389&0.2320 &0.3345&0.1139 &0.2224&\underline{0.1074} &\underline{0.2146} \\
        \midrule[0.5pt]
         \multirow{1}{*}{ETTm{1}}  &\textbf{0.0412} &\textbf{0.1245}&0.0496 &0.1382&0.0489 &0.136&0.0458 &0.1366&0.0659 &0.1682&\underline{0.045} &0.1357&0.0453 &\underline{0.1349}\\
        \midrule[0.5pt]

        \multirow{1}{*}{Illness}  &\textbf{0.1493} &\textbf{0.2188}&\underline{0.3006} &0.3367&0.3081 &0.3409&0.852 &0.5899&0.7842 &0.6832 &0.4092 &0.3739&0.3688 &\underline{0.3363}\\
        \midrule[0.5pt]
        \multirow{1}{*}{Excha.}  &\textbf{0.0029} &\textbf{0.0267}&0.0186 &0.0959&0.0101 &0.0685&0.015 &0.0835&0.0080 &\underline{0.0576} &0.0094 &0.0634&\underline{0.0079} &0.0601\\

        \midrule[0.5pt]
    \end{tabular}}
\end{table*}

\begin{figure*}[h]
\setlength{\abovecaptionskip}{0.1cm}
\centerline{\includegraphics[width=0.9\linewidth]{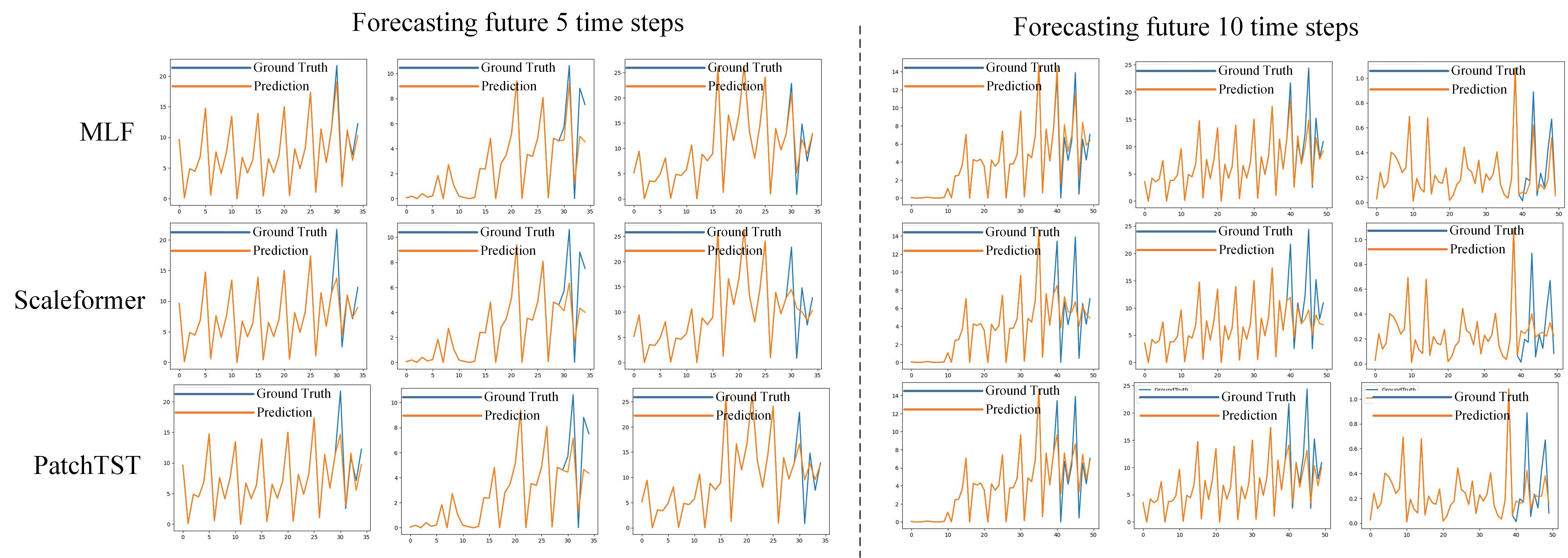} }
\caption{Visualizations of forecasting future 5 and 10 time steps on MLF and two strong baselines (Scaleformer and PtachTST) on Fund dataset.}
\label{fig:model}
\label{fig:case_vis_fund_1}
\end{figure*}

\subsection{Extra results for short-term TSF task}
Here we compare MLF with more TSF baselines:
\begin{enumerate}[noitemsep, topsep=0pt, wide=\parindent]
\item Popular transformer models Autoformer~\cite{wu2021autoformer} and FEDformer~\cite{zhou2022fedformer}.

\item Popular neural networks LSTM~\cite{hochreiter1997long}, RNN~\cite{jordan1997serial,elman1990finding}, GRU~\cite{cho2014learning} and DLinear~\cite{zeng2023transformers_linear}.

\item Popular machine learning models XGBoost~\cite{chen2015xgboost}, Prophet~\cite{taylor2018forecasting_prophet} and ARIMA~\cite{box2015time_arima}.
\end{enumerate}

\textbf{For ARIMA, }the auto-regressive order, differencing order, and moving average order are all set as 1 for better performance. \textbf{For Prophet, }the growth, changepoint range, and changepoint prior scale are set as ``linear'', 0.8 and 0.1 respectively. The seasonality prior scale and holidays prior scale are set as 10 for better performance. \textbf{For XGBoost, }the learning rate, max depth, and the number of estimators are set as 0.3, 12, and 1000 respectively. We use time and variable lag information to construct features for better performance. \textbf{For LSTM, RNN, and GRU}, the number of hidden layers is 3. The hidden size is set as 1024. 
For \textbf{Autoformer} and \textbf{FEDformer}, as with the baseline approach in the main experiments, we use the recommended hyperparameters in baseline codes.

The comparison results on Fund dataset and public datasets are shown in Table~\ref{tab:metric_fund_appendix} and Table~\ref{tab:metric_public_appendix} respectively. 
Results show that the overall accuracy of MLF is the best and second is RNN. 
The traditional machine learning models have not gained an advantage in predicting fund sales.
Moreover, we observed that models that perform well in long-term TSF task (e.g., DLinear, Autoformer, and FEDformer) do not perform as well in short-term TSF task, while MLF achieves desirable results in both short-term and long-term TSF tasks. 
This indicates the superiority of MLF.

\subsection{Extra results for patch squeeze module}
Table~\ref{tab:metric_public_squeeze_appendix} shows the effectiveness of the patch squeeze module on the ETTm1 dataset. 
We also have applied Patch Squeeze to PatchTST with fixed input length 1536. Specifically, on weather dataset, the average error of original PatchTST is 0.2554, with training epoch time 248.87s. while setting squeeze factor as 2 or 4, corresponding results are 0.2550 or 0.260 and 83.50s or 44.26s, indicating remarkable improvements in efficiency. The same conclusion is reached on the ETTm datasets. This further demonstrates that the simple but effective Patch Squeeze module significantly improves efficiency while maintaining good accuracy.

\subsection{Visualizations of forecasting on Fund dataset}
As shown in Figure~\ref{fig:case_vis_fund_1}, we visualize the forecasting of MLF and two strong baselines (Scaleformer and PatchTST) on the Fund dataset. The visualization shows that MLF can better forecast the steeper ascending and descending trends, indicating the superiority of MLF, which utilizes customized designs for addressing and using multi-period inputs for better TSF effect.

\begin{table}[bt]
    \setlength{\tabcolsep}{4pt}
    \centering
    \caption{Extra results about the effectiveness of patch squeeze module: MLF's accuracy with squeeze factors 2 (MLF-2), 4 (MLF-4), and 8 (MLF-8).}
    \label{tab:metric_public_squeeze_appendix}
    { \small
    \begin{tabular}{c|c|cc|cc|cc|cc} 
        \hline
        \multirow{2}{*}{\shortstack{}} & &  \multicolumn{2}{c|}{PatchTST} &  \multicolumn{2}{c|}{MLF-2}&  \multicolumn{2}{c|}{MLF-4}& \multicolumn{2}{c}{MLF-8}\\
         & & MSE & MAE  & MSE & MAE  & MSE & MAE  & MSE & MAE\\ 
         \midrule[0.5pt]

        \multirow{4}{*}{\rotatebox[origin=c]{90}{ETTm2}} &96 &0.166 &0.256 &0.164 &0.254 &\textbf{0.163} &\underline{0.253} &\underline{0.164} &\textbf{0.252}\\
        
         &192 &0.223 &0.296 &0.220 &\underline{0.295} &\textbf{0.218} &\textbf{0.295} &0.223 &0.297\\
         
         & 336 &0.277 &0.336 &\underline{0.275} &0.330 &\textbf{0.269} &\textbf{0.329} &0.270 &0.330\\
         
         & 720 &0.357 &0.381 &\textbf{0.343} &\underline{0.380} &\underline{0.346} &\textbf{0.378} &0.344 &0.379 \\
        \midrule[0.5pt]

    \end{tabular}
    } 
\end{table}

\end{document}